\documentclass[12pt,preprint]{aastex}
\usepackage{graphicx}
\usepackage{ulem}
\usepackage{rotating}
\usepackage{lscape}
\usepackage{natbib}
\begin{document}

\title{Disk Masses at the end of the main accretion phase:  CARMA
  Observations and Multi-Wavelength Modeling of Class I Protostars}
\author{J.A. Eisner}
\affil{Steward Observatory, The University of Arizona, 933 N. Cherry Ave, Tucson, AZ, 85721}
\email{jeisner@email.arizona.edu}


\begin{abstract}
We present imaging observations at 1.3 mm wavelength of Class I
protostars in the Taurus star forming region, obtained with the CARMA
interferometer.  Of an initial sample of 10 objects, we detected and
imaged millimeter wavelength emission from 9.  One of the 9 is
resolved into two sources, and detailed analysis of this
binary protostellar system is deferred to a future paper.  For the
remaining 8 objects, we use the CARMA data to
determine the basic morphology of the millimeter emission.  Combining
the millimeter data with 0.9 $\mu$m images of scattered light, Spitzer
IRS spectra, and broadband SEDs (all from the literature), we attempt
to determine the structure of the circumstellar material.  We consider
models including both circumstellar disks and envelopes, and constrain
the masses (and other structural parameters) of each of these
components.  We show that the disk masses in our sample span a range
from $\la 0.01$ to $\ga 0.1$ M$_{\odot}$.  The disk masses 
for our sample are significantly higher than for samples of more
evolved Class II objects.  Thus, Class I disk masses probably provide
a more accurate estimate of the initial mass budget for star and
planet formation.  However, the disk masses determined here are lower
than required by theories of giant planet formation.  The masses also
appear too low for gravitational instability, which could lead to high
mass accretion rates. Even in these Class I disks, substantial
particle growth may have hidden much of the disk mass in hard-to-see
larger bodies.
\end{abstract}

\section{Introduction}
Stars form from clouds of dust and gas.  As these clouds collapse and
conserve angular momentum, matter that is not initially located near
the polar axis falls onto a rotationally-supported disk,  Most of the
material that eventually makes up the star falls initially onto the
disk, with some matter subsequently losing angular momentum and
accreting onto the protostar.   The basic picture
of star formation thus involves the free-fall collapse of material from an
extended envelope (or cloud) onto a disk,  and then viscous accretion
through the disk and onto a protostar.

At the earliest stages of collapse, one would see a roughly spherical
cloud with no embedded protostar.  These so-called starless cores
have now been identified in a number of star forming regions
\citep[e.g.,][]{SCHNEE+07,STUTZ+09}.  As collapse proceeds, a protostar and
compact disk form, still surrounded by a substantial, roughly
spherical envelope.  Matter continues to fall onto the disk, while
viscosity sources in the disk allow net outward angular momentum
transport and inward mass accretion.   In time the envelope completes
its collapse, leaving a pre-main-sequence star surrounded by a massive
disk. The disk continues to accrete onto the star, losing mass over
time, eventually leaving the central pre-main sequence object
surrounded by only tenuous circumstellar matter.

Each of these stages of protostellar evolution produces observational
diagnostics, and an empirical picture has been developed that links
evolutionary phases with properties of the emergent spectral energy
distributions  \citep{LADA87,AWB93}.   Class 0 objects show highly
obscured SEDs that peak at sub-mm wavelengths.  These are thought to be
protostars still deeply embedded in massive envelopes.  Class I
sources show SEDs that peak in the mid-IR.  Class I objects are
thought to be surrounded by massive disks embedded in
non-spherically-symmetric envelopes \citep[e.g.,][]{EISNER+05b}. In
Class II objects light from the
central object is visible, although some short-wavelength emission is
reprocessed and emitted at longer wavelengths.  These are the T Tauri
and Herbig Ae/Be stars, known to be pre-main-sequence stars surrounded
by optically thick disks \citep[e.g.,][]{BURROWS+96,MO96,CES05}.  Finally, Class
III sources are pre-main-sequence stars either devoid of circumstellar
material or surrounded by only tenuous, optically thin circumstellar
matter \citep[e.g.,][]{PADGETT+06}.

The relative lengths of the Class 0, I, and II evolutionary stages
can be estimated by counting the number of sources belonging to
different classes in individual star-forming regions.  Absolute
timescales are determined by fitting the relatively unobscured Class
II pre-main-sequence stellar photospheric properties to theoretical
evolutionary tracks \citep[e.g.,][]{WG01}. 
The lifetime of the Class 0 phase has been
estimated as $\sim 0.1$--0.2 Myr
\citep{EVANS+08,ENOCH+09}, the Class I phase as $\sim 0.2$--0.5 Myr
\citep{WLY89,GREENE+94,KH95,EVANS+08}, and the Class II phase as $\sim
2$ Myr \citep[e.g.,][]{KENYON+90,CIEZA+07}.

The mass of circumstellar disks in each of the evolutionary stages
outlined above has crucial implications for star and planet formation.
Sufficiently massive disks may become gravitationally unstable,
providing high effective viscosities and allowing rapid growth of
protostellar mass.  The disk mass at later stages is important to planet
formation, since a certain mass is required to build giant planets
\citep[e.g.,][]{WEID+77}.

During the Class 0 phase, disks are embedded in
much more massive envelopes.  High presumed infall rates \citep[$\ga
10^{-5}$ M$_{\odot}$ yr$^{-1}$; e.g.,][]{SHU77,HCK97} from envelopes to disks
likely mean that these disks are gravitationally unstable.  
At the Class I stage, envelopes are less
massive and the disks may dominate the total system mass
\citep[e.g.,][]{JORGENSEN+09}.   Disks in Class II objects
generally appear gravitationally stable, accreting material slowly
onto their central objects \citep[$\dot{M} \la 10^{-7}$ M$_{\odot}$
yr$^{-1}$; e.g.,][]{BB89,HEG95}.
Class I sources thus offer the prospect of observing disks at an
intermediate state when they may transitioning from roiling,
gravitationally unstable disks, to more stable disks that could
provide a hospitable environment for the early stages of planet
formation.

The masses of disks surrounding Class I sources have been constrained
previously using millimeter wavelength observations
\citep[e.g.,][]{HOGERHEIJDE01,OSORIO+03,WPS03,EISNER+05b,JORGENSEN+07,JORGENSEN+09}.
Because much of
the circumstellar dust around Class I objects is expected to be
optically thin, the observed millimeter flux is correlated with mass.
However knowledge of the dust opacity and temperature (as a function
of radius) is needed in order to convert flux to mass.  Moreover,
dense regions (e.g., close to the central protostars) may be optically
thick.  Radiative transfer modeling of multi-wavelength data is
required to disentangle these effects and provide unambiguous
circumstellar mass estimates.  

Historically, constraints on the circumstellar matter in Class I
sources came from modeling of SEDs 
\citep[e.g.,][]{ALS87,KCH93,ROBITAILLE+07}.  These models include flattened,
disk-like distributions of matter that are a consequence of
angular momentum conservation \citep[e.g.,][]{ULRICH76,TSC84}.
However, the SED data can often be explained with different models as
well, including edge-on flared disk models \citep[e.g.,][]{CG99}.
Imaging data, including
scattered light images \citep[e.g.,][]{WHITNEY+03a} or millimeter
images of continuum and line emission
\citep[e.g.,][]{JORGENSEN+07,JORGENSEN+09},
have also been used to constrain the circumstellar structure of Class
I sources.  

Modeling a single dataset---e.g., an SED or scattered light image---is 
subject to ambiguities. Modeling multi-wavelength
images and fluxes simultaneously provides tighter constraints on
geometry, since emission at different wavelengths arises in different
layers of circumstellar material.  Scattered light at short wavelength
traces low-density surface layers, while longer-wavelength (e.g.,
millimeter) emission arises in deeper, denser regions.  Such modeling
has been applied to a number of Class I sources, yielding 
constraints on circumstellar morphology better than afforded by
modeling any single dataset \citep{OSORIO+03,WPS03,EISNER+05b}.

Previous estimates of disk masses in Class I objects cover a
broad range.  Modeling
millimeter images together with SEDs and scattered light images, disk
masses of 0.01--1 M$_{\odot}$ (with most masses $\ga
0.1$ M$_{\odot}$) were determined for a handful of Class I objects
\citep{OSORIO+03,WPS03,EISNER+05b}.  Somewhat lower masses, spanning
$\la 0.01$ up to $\sim 0.05$ M$_{\odot}$, were inferred from radiative
transfer modeling of sub-millimeter continuum images of 10 Class I
sources \citep{JORGENSEN+09} .  Some disk mass estimates
are consistent with simulations of protostellar
evolution, which find disk masses of $\sim 0.2$ M$_{\odot}$ for Class
I objects \citep{VOROBYOV11}.  For many objects, however, estimated
disk masses are substantially lower.

For some objects, mass estimates place Class I disks near the
limit of gravitational stability.  This finding would
fit naturally into the scenario of burst-mode accretion onto Class I
protostars \citep{VB05}, and explain the 
discrepant infall rates from the envelope to the disk and from the
disk onto the star \citep[e.g.,][]{KH87,WH04}.  In a broader sense,
the implied gravitational instability provides an easily-explained
viscosity source to enable angular momentum transport in protostellar
disks.

Estimated disk masses in Class Is appear higher than the disk masses
determined for Class II objects.  The median mass of disks in nearby
star-forming regions is estimated as $\sim 0.005$ M$_{\odot}$
\citep[e.g.,][]{AW05,AW07,EISNER+08}.  These disks are about an order
of magnitude less massive than the minimum mass solar nebula
\citep[MMSN;][]{WEID+77}, suggesting that many stars in the Galaxy lack the
circumstellar mass needed to form solar systems like ours.  Higher
inferred masses in Class I disks suggest
that substantial particle growth (which depletes the observable
population of small dust grains) has occurred between the Class I and
II stages.  Planet formation may already have advanced significantly
during the first million years of disk evolution.  Given the short
lifetimes of particles with sizes of a few centimeters to a few meters
\citep{WEIDENSCHILLING77}, this would imply growth of large
planetesimals in less than $\sim 1$ Myr.

Substantial uncertainties remain in the estimated Class I disk masses,
placing the conjectures above on somewhat shaky footing.  Our aim in
this paper is to better constrain the disk masses around a relatively
large sample of Class I sources.   We will improve upon previous constraints
using high angular resolution millimeter imaging from CARMA.
We model these data together with existing data at
other wavelengths, including broadband SEDs, Spitzer IRS spectra, and
I-band images of scattered light.

\section{Observations}
\subsection{Sample Selection}
Our sample consists of Class I sources in Taurus (Table \ref{tab:obs}).
We selected objects with existing 0.9
$\micron$ scattered light imaging \citep{EISNER+05b}, Spitzer IRS
spectra \citep{FURLAN+08,FURLAN+11}, and well-sampled SEDs
\citep[e.g.,][and references therein]{ROBITAILLE+07}.    We choose
among these the targets whose previously measured fluxes at millimeter
wavelengths are $>50$ mJy \citep{MA01}.  We apply this flux cut to
facilitate detection and mapping of millimeter emission at high
angular resolution (\S \ref{sec:carma}).  However the high fluxes in
previous low resolution observations do not guarantee high fluxes
in a smaller beam, as extended emission would be resolved out.

While the flux cut applied to the sample may lead to a bias toward
higher circumstellar masses, this bias appears to be small.
The mean 1.3 mm flux density for our sample, as measured by
\citet{MA01}, is $\sim 270$
mJy.  If we exclude the very luminous IRAS 04287+1801 (also known as
L1551 IRS 5), then the mean 1.3
mm flux density is $\sim 130$ mJy.  For all of the Taurus Class I sources in 
\citet{MA01}, the mean flux density is $\sim 190$ mJy, or $\sim 140$
mJy excluding IRAS 04287+1801.  The flux distribution of our sources is
thus not significantly different than the underlying population of
Taurus protostars.

IRAS 04263+2426 (also known as GV Tau or Haro 6-10) is a
$1\rlap{.}''2$ separation binary
\citep{LH89,KORESKO02,DNC08}, clearly resolved
into two sources in our millimeter-wavelength observations.  While we
provide basic properties of the binary, we do not include this object
in our detailed analysis and modeling.  The binarity complicates the
analysis, but also provides an opportunity to study two (presumably)
coeval sources.  We plan to devote a separate publication 
to this interesting object.

\subsection{CARMA Imaging  \label{sec:carma}}
We obtained images at 1.3 mm wavelengths for our sample using the
Combined Array for Research in Millimeter-wave Astronomy (CARMA).
CARMA consists of six 10.4 m and nine 6.1 m antennas. During our
observations the antennas were in the ``C''-configuration,  which provides
baseline separations between 30 and 350 m.  The C array observations
thus provide an angular resolution of $\sim 0\rlap{.}''8$, with sensitivity
to emission on angular scales smaller than $\sim 10''$.

Some of our continuum observations were
obtained before the full correlator implementation, and used 2 GHz of
continuum bandwidth.  Later observations were taken with the full 8
GHz of continuum bandwidth. A log of observations is given in Table
\ref{tab:obs}.  

We used MIRIAD \citep{STW95} to calibrate the CARMA data.  Bandpass
calibration utilized observations of bright, (nearly) spectrally
flat sources (see Table \ref{tab:obs}).  
To calibrate time-dependent instrumental and atmospheric effects, we
observed the blazar 3C 111 every 20 minutes.  In between 3C 111
observations, we obtained data on two or three targets.   Because our
targets are bright enough to be detected in less than a
full ``track'' (i.e., the length of time the source is above a
reasonable elevation angle), we observed multiple targets in each
track.  Targets in a single track shared calibrators, and comparison
of calibrated data for these targets provides a consistency check on
the calibration procedure.
In a few of the
tracks, Uranus was observed and used to set the instrumental flux
scale.  For the other tracks we took the flux of the gain calibrator,
3C 111, directly from the SMA submillimeter calibrators
page\footnote{http://sma1.sma.hawaii.edu/callist/callist.html}, 
extrapolating to our observing date when
necessary.

Baselines with abnormally noisy phases or high system temperatures
were deleted.  In many cases, the deleted baselines included the longest
ones, and hence the angular resolution of our observations is somewhat
less than the highest resolution possible with CARMA's C-array.  The
longest useable baselines in our data are typically $<250$ m.

\begin{figure}
\plotone{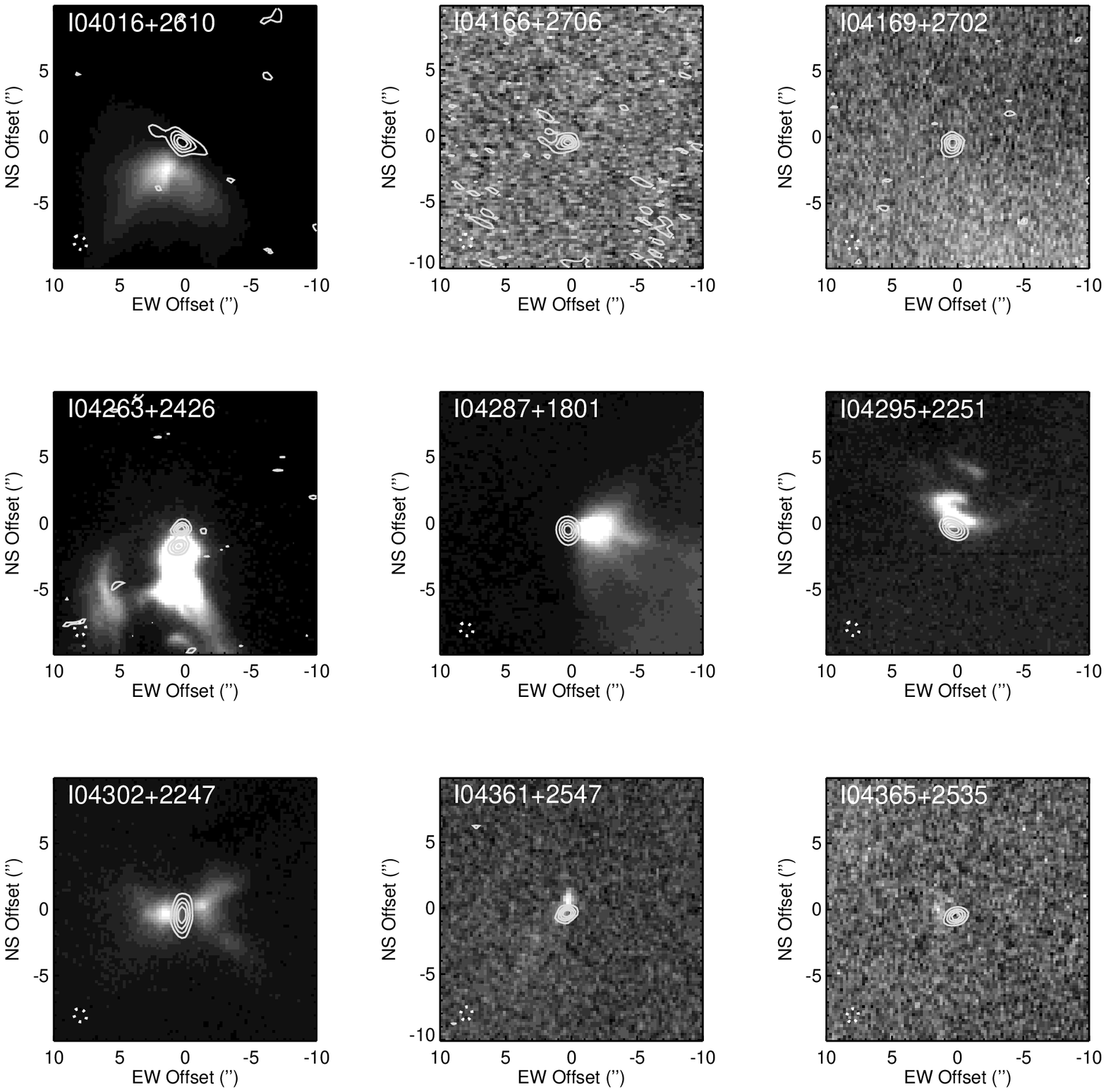}
\caption{I-band scattered-light images in grayscale with millimeter
  images overplotted as contours.  Contours are plotted in intervals
  of 20\% of the peak flux.   The dashed
  ellipse at the bottom left is the FWHM of the synthesized
  beams of the CARMA observations.}
\label{fig:nirmm}
\end{figure}

After passband, flux, and gain calibration, we INVERTed
the Fourier data and created images.  We also CLEANed the images,
determined the correct PSF for the hybrid (6-meter and 10-meter
dishes) CARMA array, 
and restored the final images using this PSF.
For one of our targets, IRAS 04181+2654, no emission was detected in
the map.  For all other targets, the mapped emission is shown in
contour plots in Figure \ref{fig:nirmm}.  Properties of the millimeter images,
including peak fluxes, 1$\sigma$ noise levels, absolute coordinates,
and PSF FWHM dimensions, are
listed in Table \ref{tab:mmims}.

While the images are useful for visualization, the visibility data
themselves are used in our analysis.    Because they are the actual
data, their uncertainties are better characterized than those of
Fourier transformed, CLEANed images. We average the visibility data
on a two-dimensional grid of $u$ and $v$ for subsequent comparison with
synthetic visibility data from models.  We compute weighted averages
in each bin, using the weights output by MIRIAD.  

The error in each
bin is given by the weighted standard deviation in the mean, scaled by
a factor that produces reasonable overall error bars.  We select scale
factors so that the error bars are similar in magnitude to the
standard deviations in the means of unweighted data.  For our data,
we use a scale factor of five for all objects except IRAS 04287+1801,
where we use a scale factor of two.

\begin{figure}
\plotone{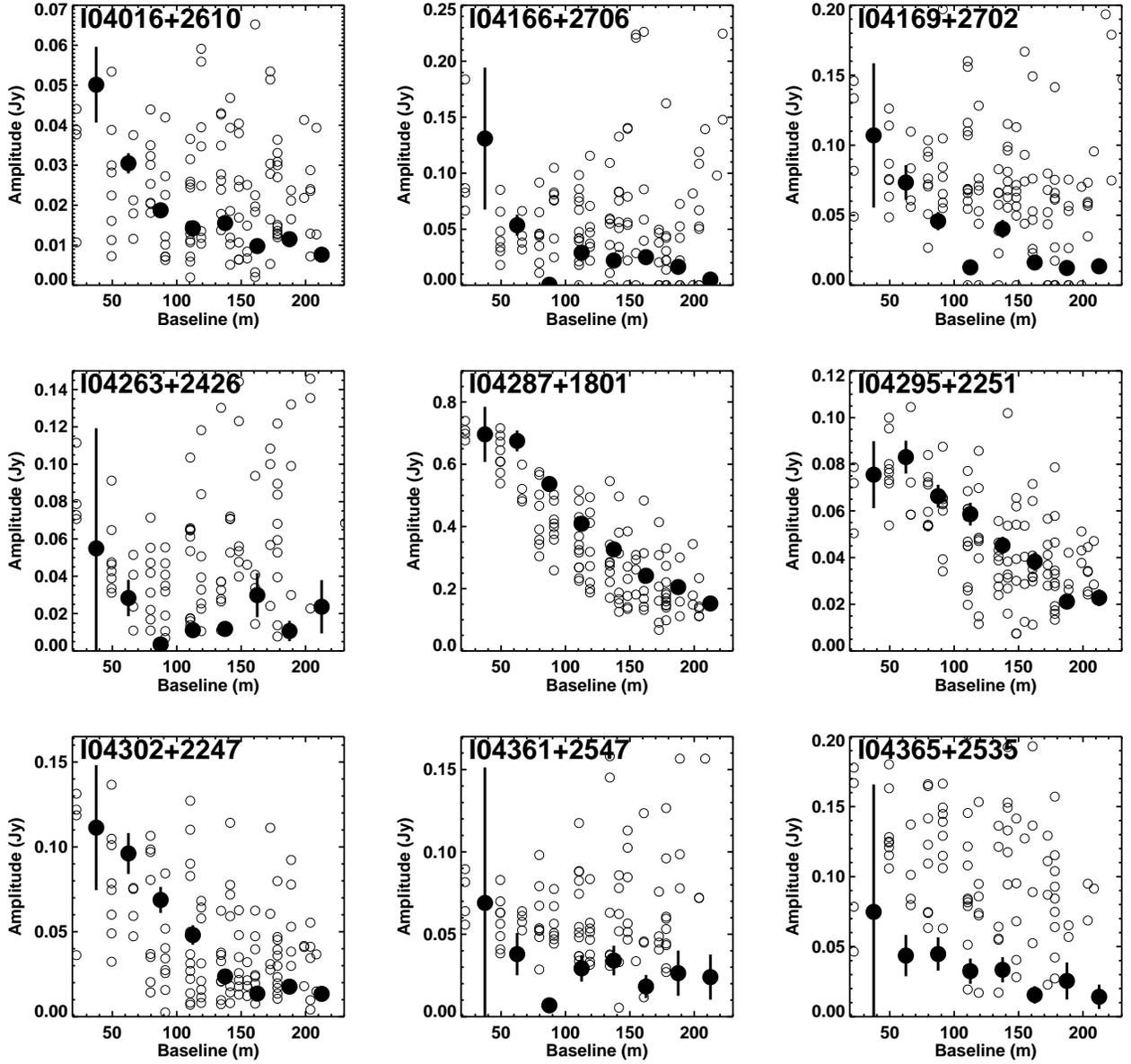}
\caption{Averaged visibility amplitudes for our targets.  Visibilities
  computed on a two-dimensional grid are shown with open circles.  The
  error bars on these points are typically fairly large (often
  comparable to the plotted range) and are not shown here.  We also
  plot the azimuthally averaged visibilities with solid points.  The
  additional averaging leads to substantially smaller uncertainties,
  indicated here with plotted error bars.  Note that the averaging is
  done coherently, and phase noise may lead to average amplitudes
  below the amplitudes measured for unaveraged data.
\label{fig:vis}}
\end{figure}

Averaged visibilities for our targets are shown in Figure
\ref{fig:vis}.  The visibilities computed by averaging over a
two-dimensional grid of $u$ and $v$ are generally quite noisy.
Azimuthally averaging these visibilities lowers the noise
substantially and allows radial structure to be more easily
discerned.  We therefore plot the azimuthally averaged visibility
amplitudes as well.

We fit inclined uniform disk models \citep[e.g.,][]{EISNER+04} to the
two-dimensional CARMA visibilities, to provide rough constraints on
the geometry of the millimeter emission.  We did not fit a uniform
disk model to IRAS 04263+2426 because it is a binary.
The fitted fluxes, sizes,
inclinations, and position angles of these disk models are listed in
Table \ref{tab:udfits}.  These fits show that the emission from most
of our sample is resolved well in our observations.  However, IRAS
04166+2706 appears to be unresolved, and the data for IRAS 04169+2702
and IRAS 04361+2547 lack the signal-to-noise and $uv$ coverage needed
to constrain the geometry well.

\subsection{Scattered Light Images  \label{sec:scatt}}

In our modeling we use Keck $I$-band imaging of our
sample, published previously in \cite{EISNER+05b}.  We calibrate the
astrometry of these images by registering to the 2MASS coordinate
system.  In each of our $I$-band images we
searched for stars also seen in 2MASS images.
We then computed a plate solution for our $I$-band images.  
The residuals in these fits are typically smaller
than $0\rlap{.}''3$, which is comparable to the error in the 2MASS
astrometry.  For the angular resolution and signal-to-noise of our
millimeter observations, we expect positional uncertainties of $\la
0\rlap{.}''1$ for the millimeter images.  The registration of $I$-band
and millimeter images is thus uncertain by $\la 0\rlap{.}''5$.

For IRAS 04263+2426 we were unable to find enough reference stars
to compute a plate solution.  This dearth of reference stars is
caused, in part, by the shorter integration time on this bright
$I$-band object (60 seconds, compared to 120--300 seconds for the other
objects in our sample).  For this source we attempted to line up the
scattered light structure seen in $J$-band 2MASS imaging with the
structure seen in the $I$-band data.  We estimate uncertainties of
$\sim 1''$ in this procedure.  

The $I$-band images, registered to the 2MASS coordinate system, are
plotted in Figure \ref{fig:nirmm}.  The scattered light morphologies for
our sample vary widely, from bright, extended emission for IRAS
04016+2610, IRAS 04263+2426, IRAS 04287+1801, IRAS 04295+2251, and
IRAS 04302+2247; to weak, more compact emission for IRAS 04361+2547
and IRAS 04365+2535; to a lack of any detectable scattered light
towards IRAS 04166+2706 and IRAS 04169+2702.

\subsection{Spectral Energy Distributions \label{sec:seds}}
We compiled broadband spectral energy distributions, including
Spitzer IRS spectra \citep{FURLAN+08,FURLAN+11}, 
from the literature
\citep{MYERS+87,WJ92,BK92,KH95,PADGETT+99,CR00,YOUNG+03,
MORIARTY-SCHIEVEN+94,HS00,MA01,AW05,KESSLER-SILACCI+05,EISNER+05b,
HARTMANN+05,ROBITAILLE+07,REBULL+10}.
We ignore the error bars quoted for these photometric measurements
(when available), and instead adopt a uniform uncertainty level.
We adopt 10\% errors for all photometric data.  However, this value is
somewhat arbitrary. As described in \S \ref{sec:fitting}, we specify
weights for various datasets in our modeling, and these weights
effectively overrule the normalization of the internal uncertainties
in a given dataset.

\section{Modeling}
Following previous work \citep[e.g.,][]{EISNER+05b},
we consider a model that includes a rotating, collapsing envelope with
an embedded, hydrostatic, Keplerian disk.  We describe the density
distribution of this model in \S \ref{sec:densdist}.  
We use the two-dimensional Monte Carlo radiative transfer code RADMC
\citep{DD04b} to compute the temperature distribution of our models.
With the density and temperature distributions computed, we use RADMC
to produce 0.9 $\mu$m images, 1.3 mm images, and SEDs spanning
wavelengths from 0.1 $\mu$m to 1 cm.  We convert the 1.3 mm images 
to two-dimensional visibilities over the same grid used to average the
CARMA visibility data (\S \ref{sec:carma}).   For ease of plotting, we
also azimuthally average these 2D visibilities.

In all of our models we fix the dust opacities.  We use dust
opacities that were computed to fit extinction measurements in dense
cores observed as part of the C2D project \citep[Pontoppidan, private
communication; see also][]{SHIRLEY+11}.  These opacities are computed
for spherical dust grains with a size distribution described by
\citet{WD01}.  The dust is a mixture of approximately 30\% amorphous
Carbon, with grains between $6 \times 10^{-4}$ and 7 $\mu$m, and 70\%
Silicate, with grains between $6 \times 10^{-4}$ and 0.6 $\mu$m.  The
grains are covered with ice mantles (70\% H$_2$O, 30\% CO$_2$), 
with an assumed ice abundance of
$3 \times 10^{-4}$ per H$_2$ molecule.
Changing dust opacities would affect the modeling,
but we lack the computational power to vary these in concert with
other parameters describing the stellar and circumstellar properties.

We consider three values of source luminosity in our analysis:
1, 3, and 10 L$_{\odot}$.  Note that these are rough approximations of
the true luminosities of our targets, which may lie between sampled
values, but computational resources limit us to a few values.
We take the stellar mass and temperature to be fixed in our modeling:
$M_{\ast}$=0.5 M$_{\sun}$ and $T_{\ast}$=4000 K.    These values are
compatible with dynamical mass estimates and effective temperature
measurements \citep[e.g.,][]{HOGERHEIJDE01,WH04}.  Thus, we are
assuming in our modeling that the only stellar parameter that varies
is the radius.

A range of viewing geometries are considered.  We allow models to be
inclined between 0 and 90$^{\circ}$ with respect to the plane of the
sky, and to have position angles spanning the full range from 0 to
360$^{\circ}$.  The position angle is defined east-of-north.

In \S \ref{sec:parstudy}, we present a
parameter study that shows how synthetic data are affected by changes
in various parameters.  Varied parameters include properties of the
circumstellar mass distribution (\S \ref{sec:densdist}),  protostellar
luminosity, and viewing geometry.
In \S \ref{sec:fitting}, we describe the grid of models we computed,
and discuss the procedure by which the best-fit model is found.

\subsection{Input Density Distribution and Stellar Parameters
 \label{sec:densdist}}
Our model includes both a collapsing envelope and an embedded disk.
The density distributions of the two components are computed
separately and then added together to produce the final model.  We
describe each of these components in turn, and demonstrate the effects
of changing input parameters on synthetic data.

\subsubsection{Collapsing Envelope \label{sec:env}}
When a rotating, initially spherical cloud of material collapses,
angular momentum conservation leads to a non-spherically symmetric
distribution. Following \cite{ULRICH76}, the density distribution of a
rotating, infalling envelope is
\begin{equation}
\rho_{\rm
  env}(r,\theta)=\frac{\dot{M}}{4\pi}(GM_{\star}r^{3})^{-1/2}\bigg(1+\frac{\mu}{\mu_{0}}
\bigg)^{-1/2}\bigg(\frac{\mu}{\mu_{0}}+2\mu_{0}^{2}\frac{R_{\rm c}}{r}\bigg)^{-1}.
\label{eq:rhoenv}
\end{equation}
Here $r$ is the stellocentric radius, $\dot{M}$ is the mass infall
rate, $M_{\star}$ is the stellar mass, $\mu=cos(\theta)$, 
$\theta$ is the angle from the polar axis, $\mu_{0}=cos(\theta_{0})$,
and $\theta_{0}$ is the initial angle of infalling material.  The
centrifugal radius, $R_{\rm c}$, is the location where material becomes
rotationally supported against gravity and the density structure
becomes less spherically distributed and more flattened:
\begin{equation}
R_c = \frac{R_0 \Omega^2}{GM_{\rm env}}.
\label{eq:rc}
\end{equation}
Here $R_0$ is the initial radius of the infalling matter, $\Omega$ is
the angular velocity, and $M_{\rm env}$ is the envelope mass.  In our modeling,
we will leave the centrifugal radius as a free parameter.

For each point in the envelope, matter can only have arrived from a
specific initial location.  We can therefore solve for $\mu_{0}$.
The particle trajectories obey the relation \citep[see, e.g.,][]{ULRICH76}
\begin{equation}\label{rrceqn}
\frac{r}{R_{c}}=\frac{\cos\theta_{0} \sin\theta_{0}^{2}}{\cos\theta_{0} -
  \cos\theta}=\frac{1-\mu_{0}^{2}}{1-\mu/\mu_{0}}.
\end{equation}
There are three solutions to this equation.  We require that the
solution be real and that $\sin \theta_{0}$ and $\sin \theta$ have the
same sign.  This means that a particle that starts out in one quadrant
of the model is required to remain in that quadrant.  The analytic
solution is:
\begin{equation}\label{muosol}
\mu_{0}=\xi-\frac{(r-R_{c})}{3R_{c}\xi},
\end{equation}
where $\xi$ is given by
\begin{equation}
\xi=\frac{\bigg[27\mu
  rR_{c}^{2}+\sqrt{729\mu^{2}r^{2}R_{c}^{4}+108R_{c}^{3}(r-R_{c})^{3}}\bigg]^{1/3}}{2^{1/3}3R_{c}}.
\end{equation}

We modify this envelope model to include an outflow cavity along the
rotation axis.  We specify this cavity as 
\begin{equation}
\rho_{\rm cav}(r,z> z_{0} + r^{\zeta}) = \rho_{\rm env} \times f_{\rm  cav}.
\label{eq:cav}
\end{equation}
Here $z$ is the height above the midplane, $r$ is the stellocentric
radius in the midplane, $z_{0}$ is the height above the midplane where
the cavity begins, and $\zeta$ describes the shape and opening angle
of the cavity.  For simplicity we fix $z_0 = 1$ AU in our models, but
we explore the effects of $\zeta$ and $f_{\rm cav}$ on synthetic data below.

The total envelope mass, $M_{\rm env}$, is obtained from the integral
of $\rho_{\rm env}$.  Since $M_{\rm env} \propto \rho_{\rm env}
\propto \dot{M}$ (Equation \ref{eq:rhoenv}),  we may use the total
envelope mass as a free parameter instead of $\dot{M}$.
The envelope mass is more relevant for our purposes, since it allows a
more direct comparison with the disk mass, $M_{\rm disk}$.

\subsubsection{Flared Disk \label{sec:disk}}
The flared disk model was first described by \cite{SS73}, where the
density distribution is given by
\begin{equation}
\rho_{disk} (r,z)=\rho_{0}\bigg(\frac{r}{r_{0}}\bigg)^{-\alpha}
\textrm{exp}\bigg(-\frac{1}{2}\bigg[\frac{z}{h(r)}\bigg]^{2}\bigg).
\label{eq:disk}
\end{equation}
Here $r$ is the radial distance from the star, $z$ is the vertical
distance above the midplane, and $r_{0}$ is assumed to be 1 AU.  The
scale height above the disk is given by 
\begin{equation}
h(r) = h_{\rm 1 AU} \bigg(\frac{r}{1 \: {\rm AU}}\bigg)^{\beta}.
\end{equation}
We adopt $\beta$=58/45, which is expected for a flared disk in
hydrostatic equilibrium  \citep[e.g.,][]{CG97}.  This
allows us to also fix the exponent of the radial volume density
profile to $\alpha=3(\beta-\frac{1}{2})=$2.37 based on viscous
accretion theory \citep{SS73}.  Note that
this implies that the surface density, $\Sigma(r) \propto r^{-1.37}$.
The value of $\alpha$ is similar to assumed values in previous
modeling of circumstellar disks \citep[e.g.][]{DALESSIO+99,CG97}, and
to the inferred value for the protosolar nebula \citep{WEID+77}.
 The free parameters for the
flared disk are $M_{\rm disk}$, $R_{\rm disk}$, and
$h_{\rm 1 AU}$.  We assume that $R_{\rm disk} = R_{\rm c}$.

\subsubsection{Disk+Envelope \label{sec:de}}
To produce a disk+envelope model, we take the envelope model from \S
\ref{sec:env} and the disk model from \S \ref{sec:disk} and add their
density distributions.    In our modeling we vary the disk and
envelope masses so that our models encompass both disk-dominated and
envelope-dominated morphologies.

\subsection{Parameter Study \label{sec:parstudy}}
Here we vary the input parameters of the disk+envelope model described
above, and illustrate the effect these parameters have on synthetic
SEDs, 0.9 $\mu$m scattered light images, and millimeter wavelength
visibilities.  We vary parameters about a baseline model, which has
the following parameters: $M_{\rm disk} = 0.1$ M$_{\odot}$, $h_{\rm 1 AU} =
0.05$ AU, $R_{\rm disk} = R_{\rm c} = 30$ AU, $M_{\rm
  env}$ = 0.05 M$_{\odot}$, 
$R_{\rm out} = 1000$ AU, $f_{\rm cav} = 0.2$, $\zeta = 1.0$, 
$L=3$ L$_{\odot}$, $i$=45$^{\circ}$, and PA=$90^{\circ}$.

\subsubsection{Disk Parameters}
In our disk+envelope model we have assumed a power-law radial surface
density profile declining with radius with an exponent of -1.37.  
Since the disk surface area grows as the square of radius, the radial
mass distribution increases with radius: $M(r) \propto r^{0.63}$. 
However since the disk outer radius is equal to the envelope
centrifugal radius, material in the disk is still compact compared to
the envelope mass distribution.  Note, too, that the temperature is
higher at smaller radii, and the flux from optically thin material can
still be centrally concentrated even if the mass distribution is
weighted to larger radii.

\begin{figure}
\plotone{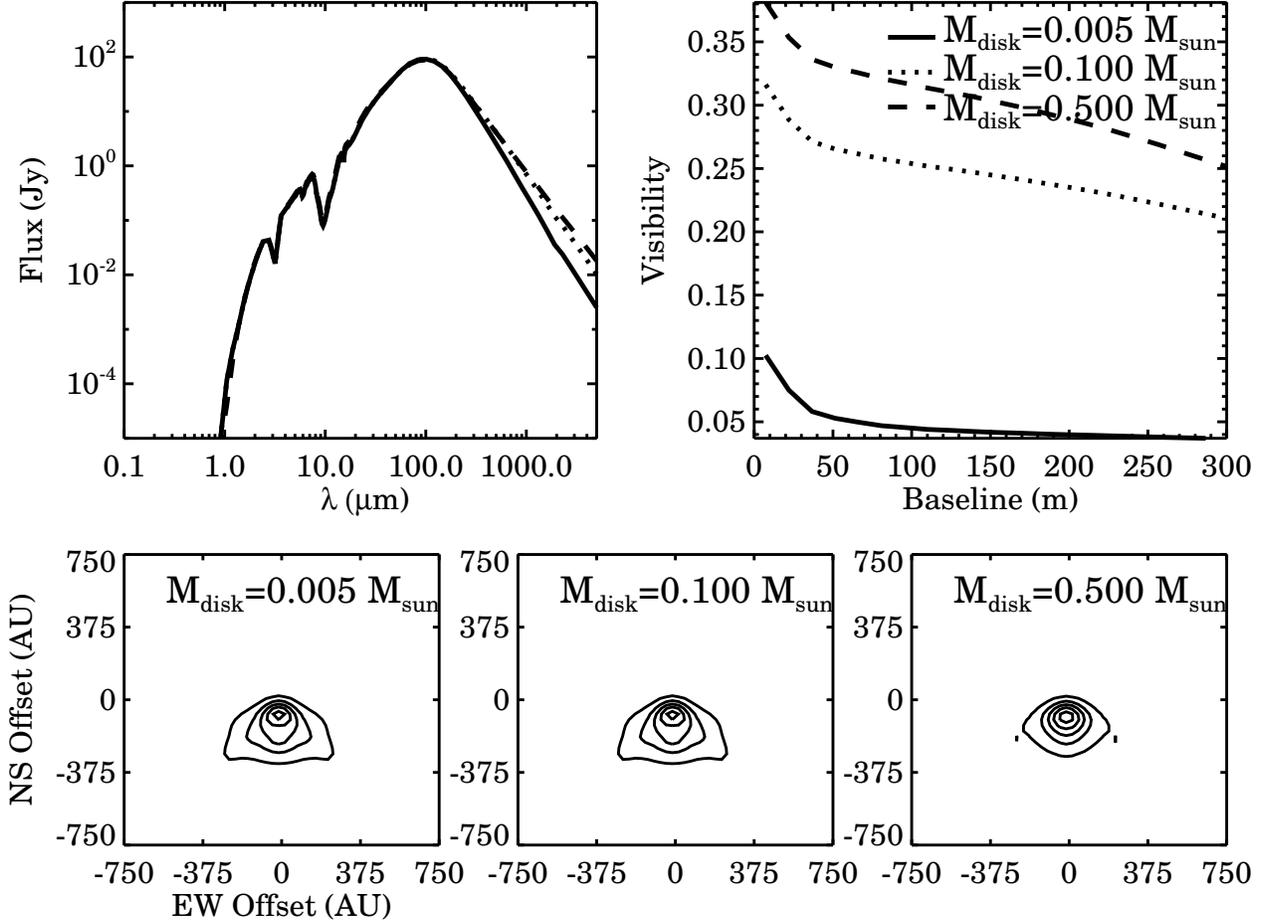}
\caption{Fluxes ({\it top left}), normalized $\lambda$1.3mm visibility
  profiles ({\it top right}), and 0.9
  $\micron$ images ({\it bottom}) for disk+envelope models with different
  disk masses.  Contours are
  plotted for 10\%, 30\%, 50\%, 70\%, and 90\% of the peak flux
  levels.  As listed in \S \ref{sec:parstudy}, 
all models have $M_{\rm env} = 0.05$ M$_{\odot}$,
  $\alpha = -2.37$, $h_{\rm 1 \: AU} = 0.05$ AU,  
$R_{\rm disk}= R_{\rm c}=30$ AU, $R_{\rm out}=1000$ AU,
  $f_{\rm cav} = 0.2$, $\zeta=1.0$, $L_{\ast} = 3$ L$_{\odot}$,
  $i=45^{\circ}$, and position angle $=90^{\circ}$.  
\label{fig:mdisk}}
\end{figure}

A higher mass corresponds to an increased mm-wavelength flux. As
seen in Figure \ref{fig:mdisk}, the long-wavelength SED as well as the
normalization of the mm-wavelength visibility is increased as the disk
mass is increased.   The disk emission is unresolved for our baseline
disk model radius ($R_{\rm disk} = 30$ AU).  However for $R_{\rm disk}
\ga 100$ AU, some disk emission may be resolved on the longest
baselines.

The effect of disk mass on the scattered light image
is small since most of the scattered light is generated on larger
scales where the envelope component dominates. Figure \ref{fig:mdisk}
shows a slightly less extended scattered light morphology for the
highest disk mass considered.  However this may reflect the fact that
fewer photons escaped the high optical depth during the radiative
transfer calculation.

\begin{figure}
\plotone{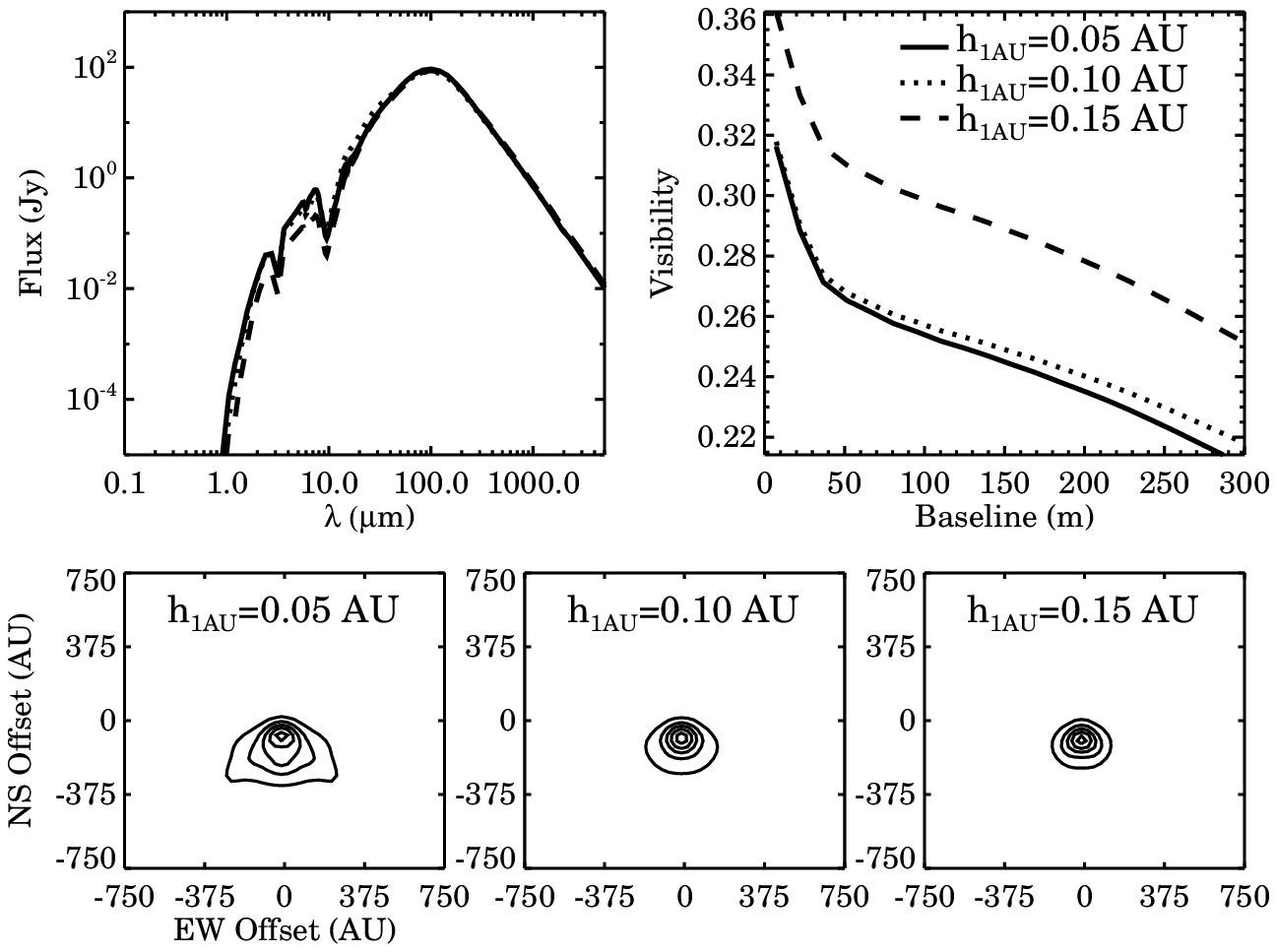}
\caption{Fluxes ({\it top left}), normalized $\lambda$1.3mm visibility
  profiles ({\it top right}), and 0.9
  $\micron$ images ({\it bottom}) for disk+envelope models with different
  disk scale heights.  Contours are
  plotted for 10\%, 30\%, 50\%, 70\%, and 90\% of the peak flux
  levels.   Parameters of the basic model are listed in \S
  \ref{sec:parstudy} and in the caption of Figure \ref{fig:mdisk}.
\label{fig:hr}}
\end{figure}

Larger disk scale heights
mean that the disk is better at intercepting short-wavelength light
and reprocessing it to longer wavelengths.  Thus models with higher
values of $h_{\rm 1 AU}$ show depressed optical/near-IR fluxes and increased millimeter
fluxes (Figure \ref{fig:hr}).  Scattered light images become more
compact for larger values of $h_{\rm 1 AU}$, because photons can be scattered
from vertically-extended disk regions at smaller radii.

\begin{figure}
\plotone{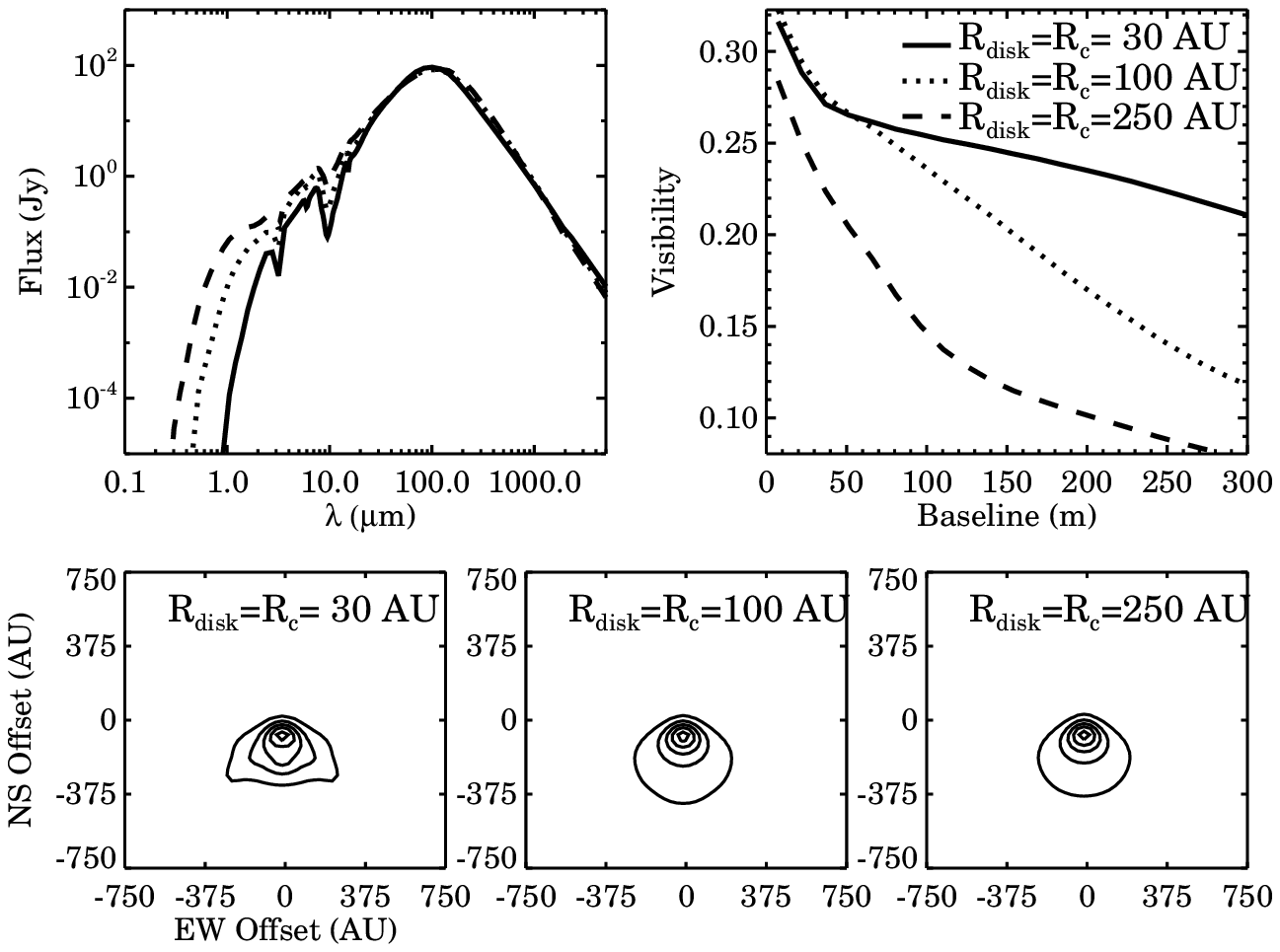}
\caption{Fluxes ({\it top left}), normalized $\lambda$1.3mm visibility
  profiles ({\it top right}), and 0.9
  $\micron$ images ({\it bottom}) for disk+envelope models with different
  disk/centrifugal radii.  Contours are
  plotted for 10\%, 30\%, 50\%, 70\%, and 90\% of the peak flux
  levels.  Parameters of the basic model are listed in \S
  \ref{sec:parstudy} and in the caption of Figure \ref{fig:mdisk}.
\label{fig:rc}}
\end{figure}

A larger disk outer radius means more resolved millimeter visibilities
(Figure \ref{fig:rc}).  Larger disks also re-distribute mass to larger
radii, thus decreasing the optical depth at small radii and allowing
more infrared emission to escape.  Because the disk outer radius is
also the envelope centrifugal radius in our model, its effects on the
synthetic data are somewhat more complicated.  We discuss how
centrifugal radius impacts the synthetic data below.

The disk surface density profile may also influence the observed radial
profile of 1 mm emission, since the mostly-optically-thin dust
emission correlates with mass.  Shallower radial surface density
profiles produce more millimeter emission at larger stellocentric
radii, and hence more resolved visibilities.  However, in the interest
of minimizing the number of free parameters, we consider the surface
density to be fixed in our modeling.

\subsubsection{Envelope Parameters}
\begin{figure}
\plotone{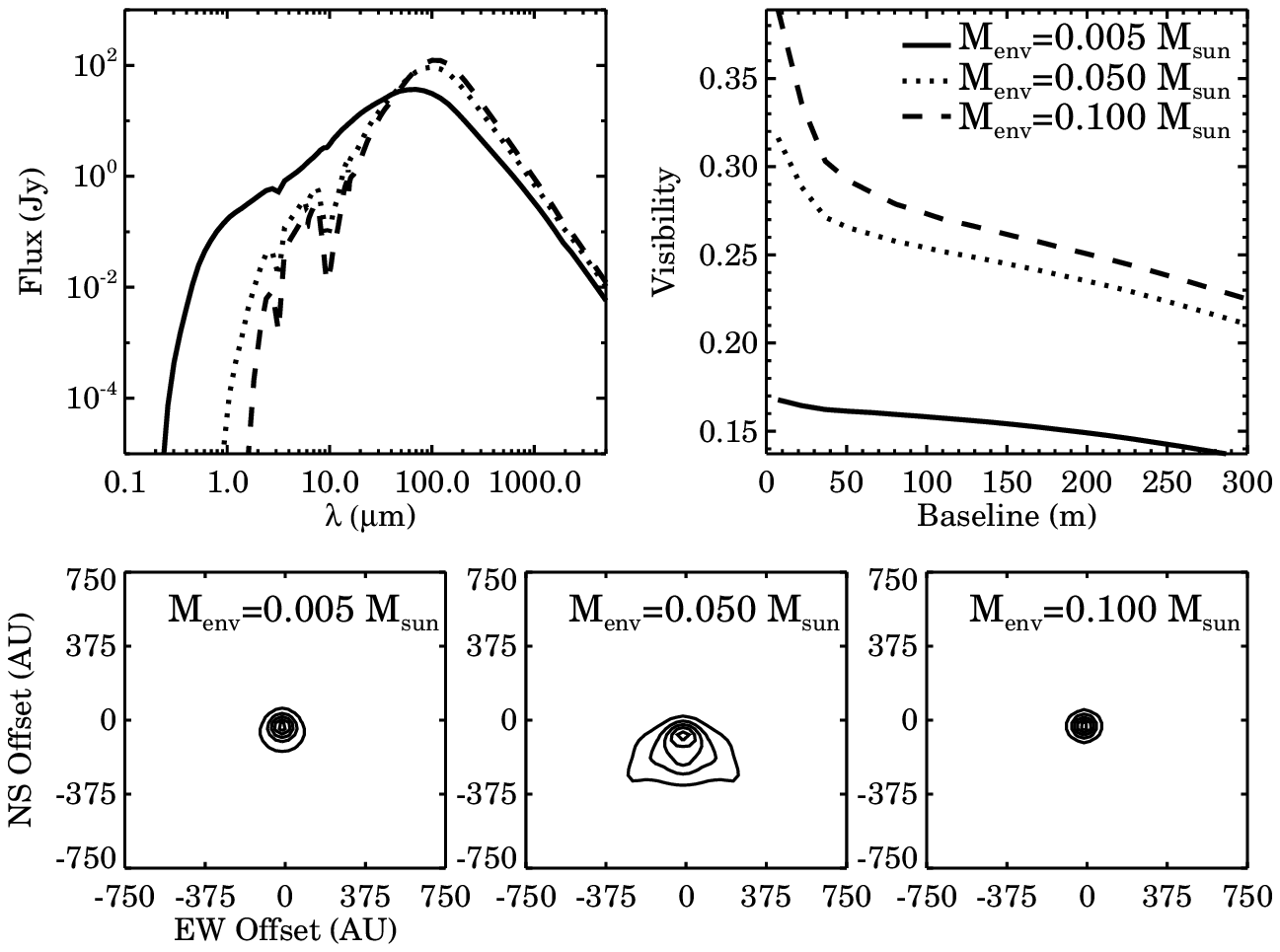}
\caption{Fluxes ({\it top left}), normalized $\lambda$1.3mm visibility
  profiles ({\it top right}), and 0.9
  $\micron$ images ({\it bottom}) for disk+envelope models with different
  envelope masses.  Contours are
  plotted for 10\%, 30\%, 50\%, 70\%, and 90\% of the peak flux
  levels.   Parameters of the basic model are listed in \S
  \ref{sec:parstudy} and in the caption of Figure \ref{fig:mdisk}.
\label{fig:menv}}
\end{figure}

Increasing the envelope mass increases the extinction toward the
central object, thus decreasing the optical and infrared fluxes, and increases the
mass of cooler material, thus increasing the millimeter fluxes (Figure
\ref{fig:menv}).  Unlike the disk mass, envelope mass is distributed
over larger spatial scales, and much of the additional millimeter
emission is resolved by the long-baseline CARMA observations.
Increasing the envelope mass thus tends to increase the visibility
amplitudes more at shorter baselines than at longer baselines.

We see in Figure \ref{fig:menv} that the effect on the synthetic
scattered light images of changing the
envelope massis not monotonic.
For lower envelope masses, the central object
(protostar+inner disk) is less obscured and the extended scattered
light is difficult to discern against the bright background.  As
envelope mass increases, the central light is blocked and it is easier
to see the extended scattered component.  For still higher envelope
masses, scattered light simply can not escape efficiently, and we see
the obscured central object at a very low level.

As discussed above, since $R_{\rm c} = R_{\rm
  disk}$, a larger value of $R_{\rm c}$  means that the disk mass is
distributed over a larger range of radii. A larger centrifugal
radius also implies that a larger fraction of the total envelope mass is
distributed in a flattened distribution.   The millimeter visibilities are thus more
resolved for larger values of $R_{\rm c}$.

Because a larger centrifugal radius means that the more spherical
component of the envelope begins at larger stellocentric radii, it
also means that I-band emission can escape more easily.  As seen in
Figure \ref{fig:rc}, the strength of the scattered light increases
with $R_{\rm c}$.   We also see changes in the morphology of the
scattered light images.  It seems that when the disk size (defined by
$R_{\rm c}$) becomes large enough, scattered light from the disk may
dominate the image.  For smaller disks, and $R_{\rm c}$ values, the
scattering may come predominantly from the envelope component.

\begin{figure}
\plotone{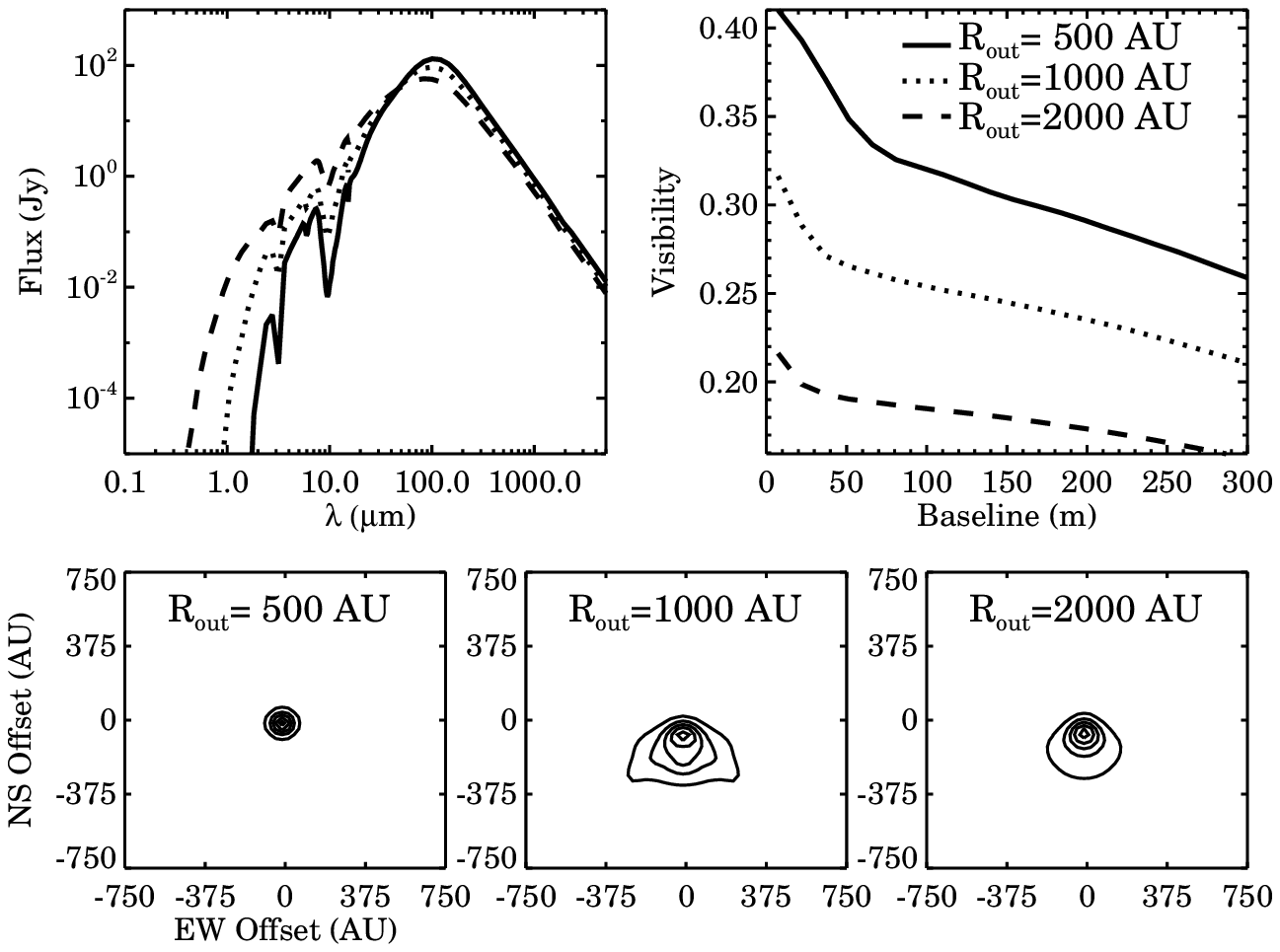}
\caption{Fluxes ({\it top left}), normalized $\lambda$1.3mm visibility
  profiles ({\it top right}), and 0.9
  $\micron$ images ({\it bottom}) for disk+envelope models with different
  envelope outer radii. Contours are
  plotted for 10\%, 30\%, 50\%, 70\%, and 90\% of the peak flux
  levels.  Parameters of the basic model are listed in \S
  \ref{sec:parstudy} and in the caption of Figure \ref{fig:mdisk}.
\label{fig:rout}}
\end{figure}

Figure \ref{fig:rout} shows the effect on synthesized data of
varying $R_{\rm out}$.   Because the envelope mass is held fixed, a
larger value of $R_{\rm out}$ means that the mass is more spread out
and the mass on small scales is decreased. Models with larger outer
radii are more resolved (i.e., display lower visibilities).  They also
show somewhat lower total millimeter fluxes, since more of the mass is
shifted to lower temperature regions.

Envelopes with larger outer radii (and hence less mass on compact
scales) let more short-wavelength flux escape to the observer.  This is
reflected clearly in the SED shown in Figure \ref{fig:rout}. For small
$R_{\rm out}$ values, little light escapes and the scattered light
image shows only highly extincted flux from the central object.  For
high values of $R_{\rm out}$, flux can escape from the less dense
inner regions of the envelope.  For still larger values of $R_{\rm
  out}$, the inner regions of the envelope become sufficiently tenuous
that the denser disk component produces most of the scattered light.
 
\subsubsection{Outflow Cavity Properties}
The outflow cavity in our models is described by three parameters:
$z_0$, $\zeta$, and $f_{\rm cav}$ (Equation \ref{eq:cav}).  $z_0$ is
not particularly important to the synthetic data and so we fix its
value.  $\zeta$ and $f_{\rm cav}$ influence both the shape of the
cavity and its contrast with respect to the rest of the envelope (and
whether there is a cavity at all).

\begin{figure}
\plotone{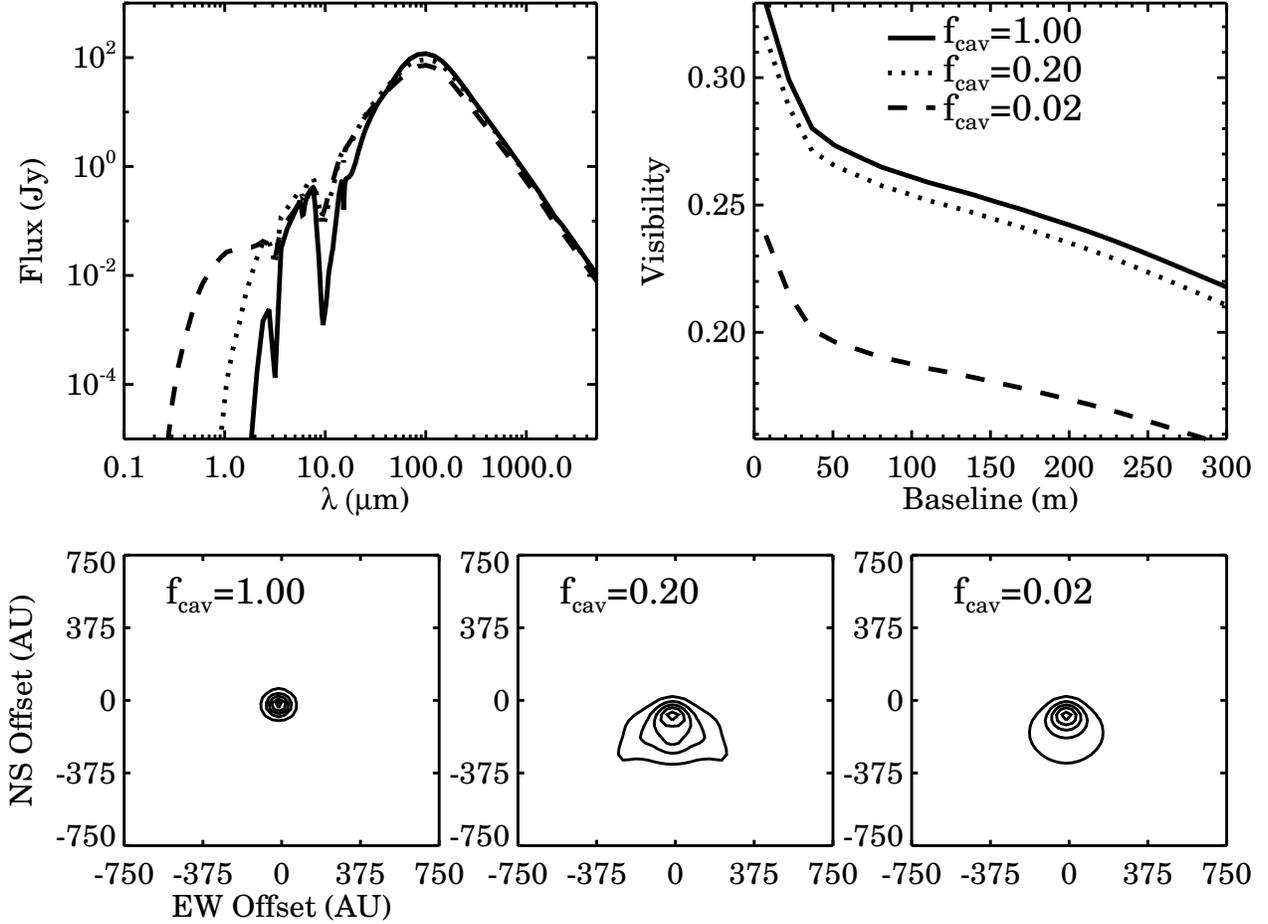}
\caption{Fluxes ({\it top left}), normalized $\lambda$1.3mm visibility
  profiles ({\it top right}), and 0.9
  $\micron$ images ({\it bottom}) for disk+envelope models with different
  density contrasts within the outflow cavity. Contours are
  plotted for 10\%, 30\%, 50\%, 70\%, and 90\% of the peak flux
  levels.  Parameters of the basic model are listed in \S
  \ref{sec:parstudy} and in the caption of Figure \ref{fig:mdisk}.
\label{fig:fcav}}
\end{figure}

A smaller value of $f_{\rm cav}$, corresponding to a higher contrast
between the densities within and outside the outflow cavity,
facilitates the escape of short-wavelength emission.  Smaller $f_{\rm
  cav}$ values lead to more compact scattered light images, since
more light from the central disk is visible.  For very large values of
$f_{\rm cav}$, very little scattered light escapes, and we see the
central source at a low level (Figure \ref{fig:fcav}).

\begin{figure}
\plotone{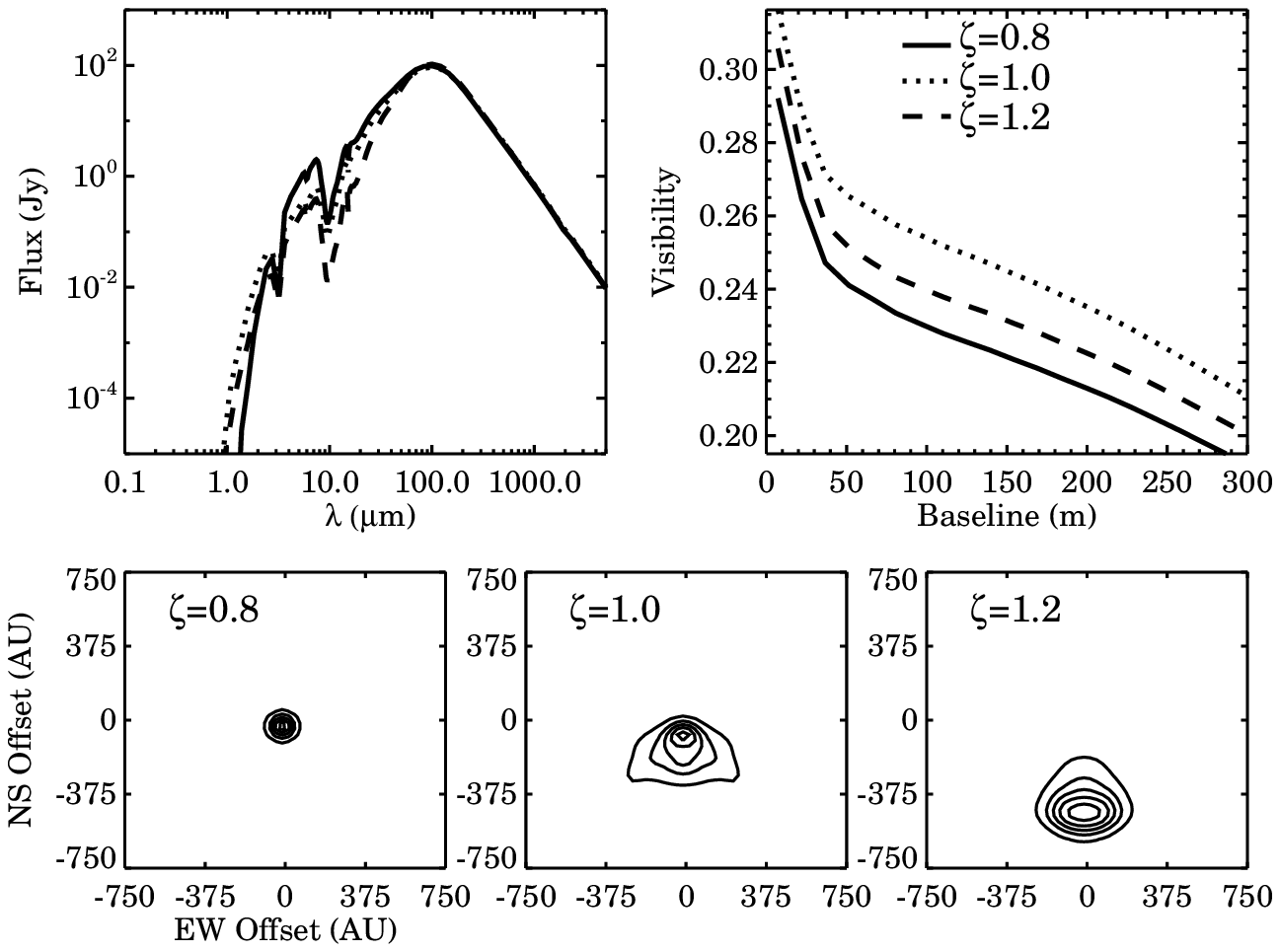}
\caption{Fluxes ({\it top left}), normalized $\lambda$1.3mm visibility
  profiles ({\it top right}), and 0.9
  $\micron$ images ({\it bottom}) for disk+envelope models with different
  outflow cavity shapes.  Contours are
  plotted for 10\%, 30\%, 50\%, 70\%, and 90\% of the peak flux
  levels.  Parameters of the basic model are listed in \S
  \ref{sec:parstudy} and in the caption of Figure \ref{fig:mdisk}.
\label{fig:zeta}}
\end{figure}

In Figure \ref{fig:zeta} we show the effects of varying $\zeta$
on the synthetic data.  While Equation \ref{eq:cav} describes a curved
outflow surface, we can equate opening angles with various values of
$\zeta$.  The considered values of $\zeta = 0.8, 1.0$, and 1.2
correspond roughly with
half-opening-angles of $70^{\circ}$, $45^{\circ}$, and $20^{\circ}$,
respectively.  Because mass is conserved, larger outflow opening
angles (smaller $\zeta$) mean more mass is distributed in a flattened
density distribution.  This facilitates the escape of short-wavelength
emission.  Larger values of $\zeta$ produce scattered light
images with larger offsets from the central object, since the observed
line of sight intersects the cavity at larger radii than for smaller
values of $\zeta$.

\subsubsection{Source Luminosity}
\begin{figure}
\plotone{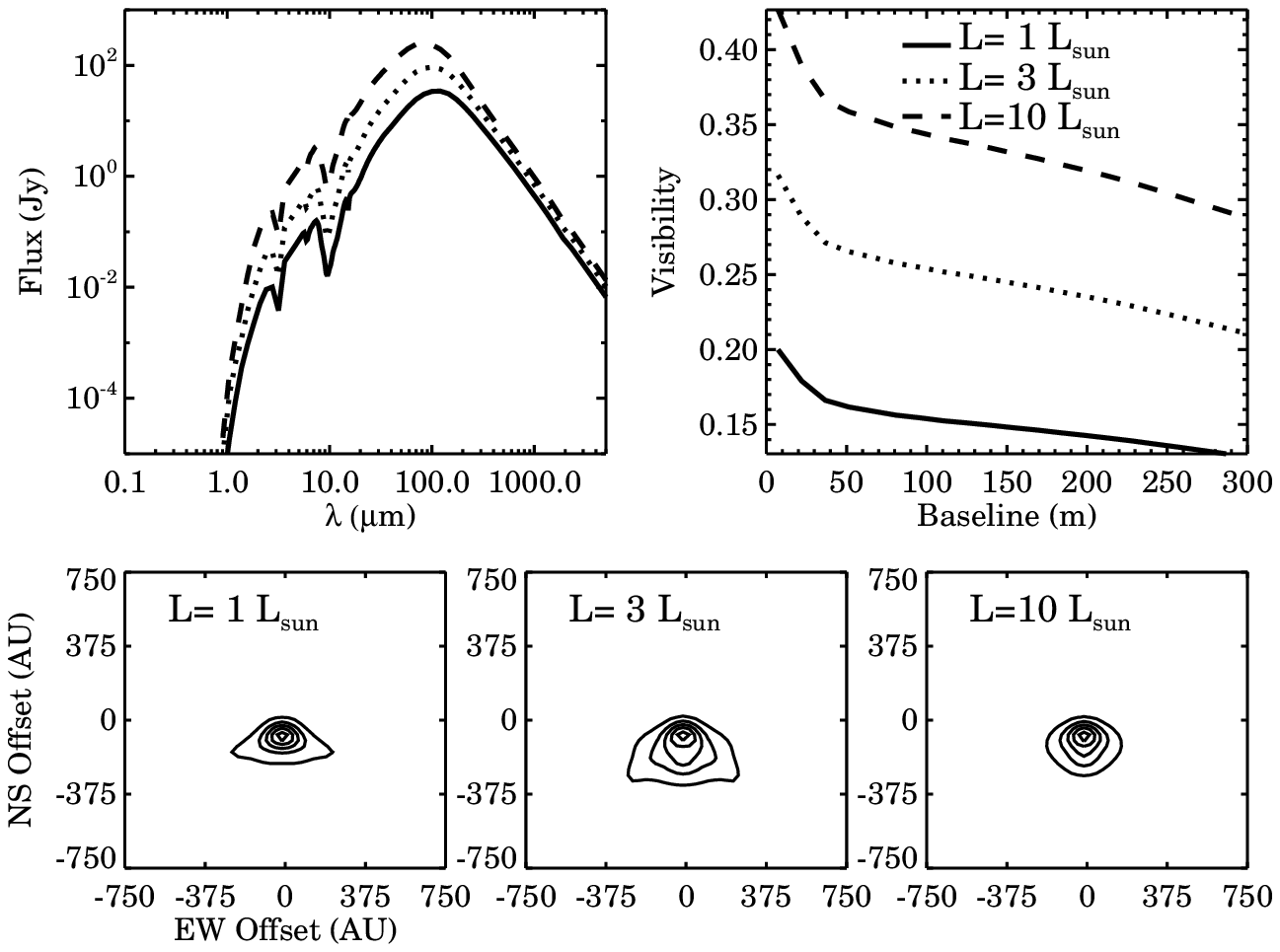}
\caption{Fluxes ({\it top left}), normalized $\lambda$1.3mm visibility
  profiles ({\it top right}), and 0.9
  $\micron$ images ({\it bottom}) for disk+envelope models with different
source luminosities.  Contours are
  plotted for 10\%, 30\%, 50\%, 70\%, and 90\% of the peak flux
  levels.  Parameters of the basic model are listed in \S
  \ref{sec:parstudy} and in the caption of Figure \ref{fig:mdisk}.
\label{fig:lum}}
\end{figure}

The central luminosity of the model strongly influences the synthetic
fluxes.  However, the fluxes do not scale simply with luminosity.  As
seen in Figure \ref{fig:lum}, the infrared fluxes increase much
more than then mm fluxes as source luminosity rises.  
The morphology of the millimeter emission, as well as
the scattered light emission, are relatively unaffected by the central
luminosity.

\subsubsection{Viewing Geometry}
\begin{figure}
\plotone{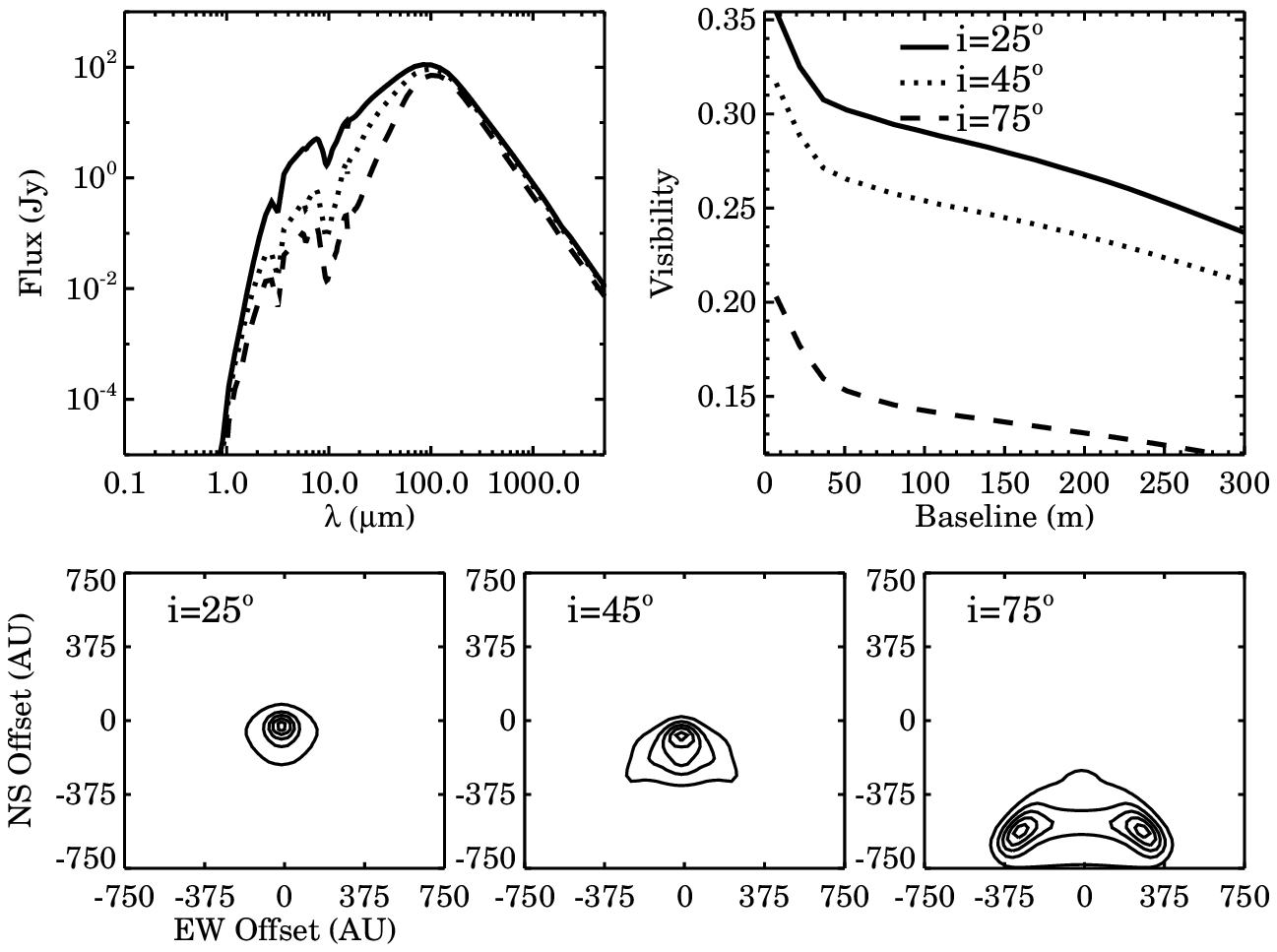}
\caption{Fluxes ({\it top left}), normalized $\lambda$1.3mm visibility
  profiles ({\it top right}), and 0.9
  $\micron$ images ({\it bottom}) for disk+envelope models with different
  viewing geometries.  
Contours are
  plotted for 10\%, 30\%, 50\%, 70\%, and 90\% of the peak flux
  levels.  Parameters of the basic model are listed in \S
  \ref{sec:parstudy} and in the caption of Figure \ref{fig:mdisk}.
\label{fig:inc}}
\end{figure}

Viewing geometry dictates whether we see the disk+envelope model through
the outflow cavity, through the dense disk midplane, or through some
intermediate region.  For more edge-on inclinations, the central
regions of the disk and envelope are more heavily obscured.  Scattered light is
thus more extended and offset from the center (Figure \ref{fig:inc}). 
While inclination does
not change the size of the millimeter emission, it does affect the
azimuthally averaged visibilities (since the minor axis becomes less
resolved for higher inclinations).  The position angle (PA) simply rotates
the scattered light images.  PA does not alter the azimuthally
averaged visibilities, but it does affect the 2D synthetic
visibilities actually used in the modeling.

\subsection{Model Fitting \label{sec:fitting}}
We generate a grid of models and 
compute the $\chi^2$ residuals between each model and the combined
dataset.  Each model takes $\gg 1$ hour to generate, and so we are
limited to a fairly small grid.  We can not afford to vary
all of the parameters discussed above simultaneously.  We have chosen
to vary $M_{\rm disk}$, $h_{\rm 1 AU}$, $R_{\rm disk} = R_{\rm c}$, $M_{\rm env}$, 
$R_{\rm out}$, $f_{\rm cav}$, $L$, $i$, and PA.  We fix $\zeta = 1.0$
and $\alpha=-2.37$.

We generated models for: $M_{\rm disk} = 0.005,0.01,0.1$,and $0.5$
M$_{\odot}$; $h_{\rm 1 AU} = 0.05$ and 0.15 AU; $R_{\rm disk} = R_{\rm
  c} = 30, 100, 250$, and 450 AU; 
$M_{\rm env}= 0.005,0.01,0.05$, and 0.1
M$_{\odot}$; $R_{\rm out} = 500,1000$, and 2000 AU; 
$f_{\rm cav} = 1$ and 5; 
and $L = 1,3$, and 10 L$_{\odot}$.  The
inclination and position angle can be varied after the temperature
distribution has been computed, and so generating a large range of
viewing geometries is relatively inexpensive.  We thus consider all
inclinations in steps of $5^{\circ}$ and all position angles in steps
of $10^{\circ}$ in our modeling.


Because the weights of
different datasets are somewhat arbitrary (for example, 
they depend on number of observed datapoints), we adjust the weights to
ensure a roughly equal weighting of the different datasets.  Following
\citet{EISNER+05b}, we divide each dataset by the minimum $\chi^2$
residual between all models and that dataset:
\begin{equation}
\chi_{\rm TOT}^{2}=\frac{\chi_{\rm mm}^{2}}{\rm min(\chi_{\rm
    mm}^{2})}+\frac{\chi_{\rm Iband}^{2}}{\rm min(\chi_{\rm
    Iband}^{2})}+\frac{\chi_{\rm SED}^{2}}{\rm min(\chi_{\rm
    SED}^{2})}.
\end{equation}
We include additional scaling factors here that we may
adjust to ensure roughly equal weights given to each dataset.
Adjusting these scale factors also enables us to illustrate the
constraints arising from individual datasets.
The reduced $\chi^2$ we adopt is given by
\begin{equation}
\chi_r^2 = \left(w_{\rm mm}\frac{\chi_{\rm mm}^{2}}{\rm min(\chi_{\rm
    mm}^{2})}+w_{\rm Iband} \frac{\chi_{\rm Iband}^{2}}{\rm min(\chi_{\rm
    Iband}^{2})}+w_{\rm SED} \frac{\chi_{\rm SED}^{2}}{\rm min(\chi_{\rm
    SED}^{2})}\right) / \left(w_{\rm mm} + w_{\rm Iband} + w_{\rm SED}\right),
\label{eq:chi2}
\end{equation}
where $w$ indicates the weighting factor for an individual dataset.

\section{Results}

Table \ref{tab:modfits} lists the best fit model parameters for each
object in our sample, and Figures \ref{fig:i04016}--\ref{fig:i04365}
show observed and synthetic data for these models. 
For each object, we varied the relative weights
of different datasets, to highlight the constraints imposed by each
set of observations.   Changing the relative weights allows us to
explore the region around the $\chi^2$ minimum, and hence to determine
how shallow
the $\chi^2$ surface is when projected onto different parameters.  For
example, if a parameter is very well constrained then it will remain
constant as data weights are varied.  In contrast, some parameters may
explore large values without significantly affecting fit quality.

In general, several parameters are well constrained, staying roughly
constant as relative weights are varied.  These include $M_{\rm
  disk}$, $L_{\ast}$, inclination, and position angle.  This probably
reflects the fact that these parameters tend to strongly influence one
dataset, and to produce an effect that does not strongly resemble the
effects of other model parameters.  The main effect of $M_{\rm disk}$
is on the shape and strength of the millimeter visibilities (Figure
\ref{fig:mdisk}).  While $L_{\ast}$ does not directly translate into a
flux normalization, its effects are mainly observed in the SED and the
normalization of the millimeter visibilities.  Since inclination and
position angle produce strong, easily discernible effects on the
synthetic scattered light images, their values can be constrained
well.

In contrast, other parameters explore a wide range of values without
significantly changing the overall fit quality.  For example, $R_{\rm
  out}$ can vary over the entire range of explored values (from 500 to
2000 AU) in some cases.  To some extent this reflects degeneracy among
different
parameters.  As seen in Table \ref{tab:modfits}, $M_{\rm env}$ varies
in concert with $R_{\rm out}$.  This is because larger $R_{\rm out}$
means that a given envelope mass is divided over a larger range of
radii.  As described in \S \ref{sec:parstudy}, other model parameters may
also produce degenerate effects on various synthetic data.

The sparse sampling of our model grid occasionally leads to
large apparent differences between models.  In particular we only
sample values of 1, 3, and 10 $L_{\odot}$ for $L_{\ast}$. Fitted
values of 3 and 10 $L_{\odot}$ for a given source (e.g., IRAS
04016+2610; Table \ref{tab:modfits}) do not necessarily
indicate a factor of 3 uncertainty in the source luminosity.   Rather,
this probably reflects that the true source luminosity lies between these
values.  This point can be generalized to all parameters in our grid,
which are sampled sparsely.

The main goals of our study are to constrain the disk mass, envelope
mass, and their ratio.  From Table \ref{tab:modfits}, values for each
of these can be extracted.   We compute the average disk mass for each
of the four best-fit models.  The median disk mass for all the objects
(using these averages) is 0.008 M$_{\odot}$.  The median envelope mass
is 0.08 M$_{\odot}$.  For the overall sample, the envelope mass is
$\sim 10$ times higher than the disk mass.

Another way to describe disk and envelope masses 
is to estimate the amount of mass on small scales relative
to the more extended mass distribution.  This circumvents uncertain
interpretations of how much of the mass in our envelope model--which
includes some material in a flattened, disk-like distribution within
$R_{\rm c}$--should be counted as part to the disk mass.

\begin{figure}
\plotone{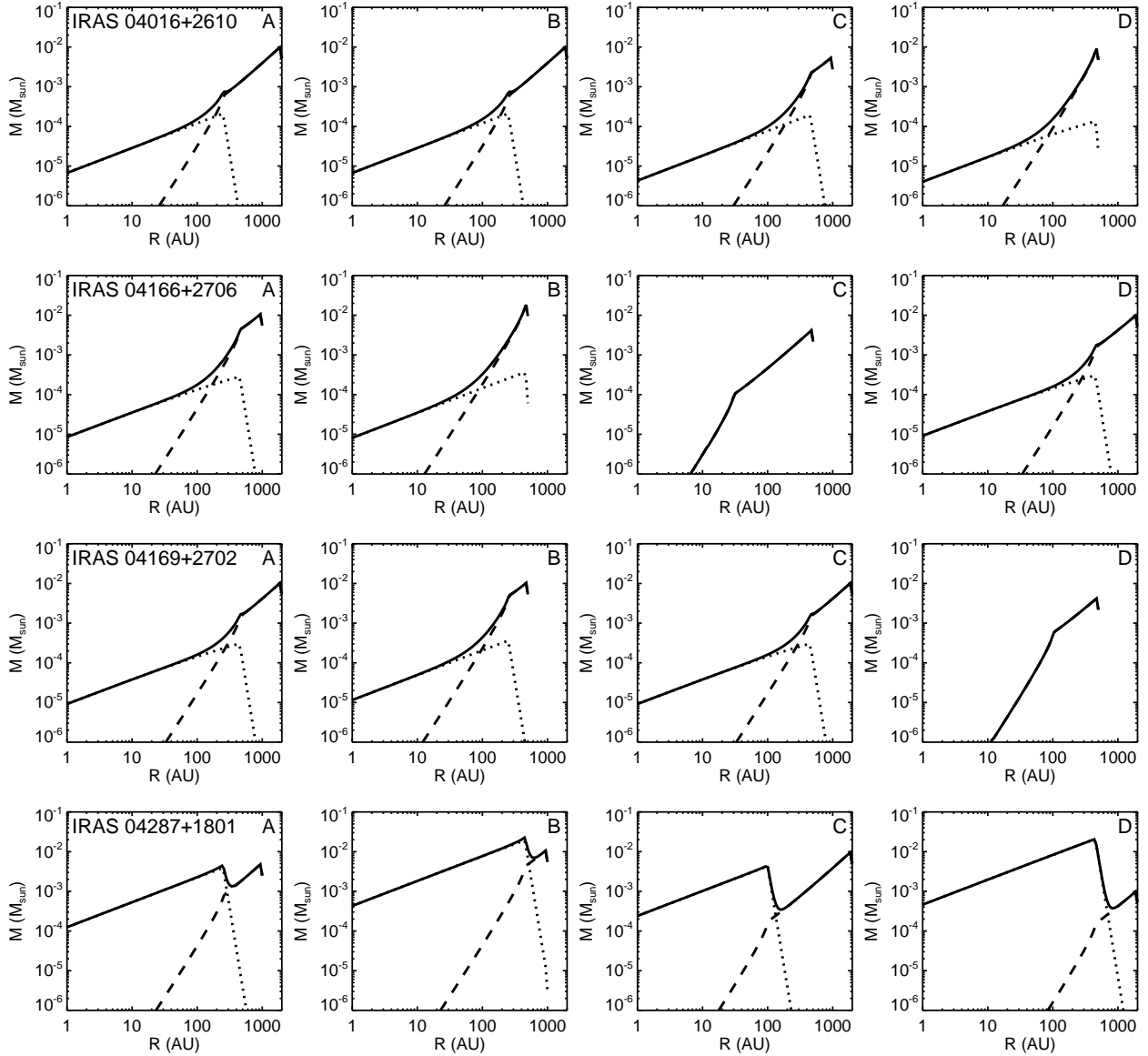}
\caption{Radial mass distributions for all the models listed in Table
  \ref{tab:modfits}.  We show the mass distributions for the disk
  components ({\it dotted curves}), the envelope components ({\it
    dashed curves}), and the total disk+envelope models ({\it solid
    curves}).  Each row show the mass profiles for model fits to a single
  object, with the ordering of columns (left-to-right) 
corresponding to the ordering of rows (top-to-bottom)
  in Table \ref{tab:modfits}.
\label{fig:mprofiles}}
\end{figure}

In Figure \ref{fig:mprofiles} we plot the azimuthally-averaged mass
distributions for all of the models listed in Table
\ref{tab:modfits}.  It is evident that for most objects in our sample,
the majority of the mass is found at larger radii.  This is true even
for the disk component, since the assumed surface density profile
($\Sigma \propto r^{-1.37}$) implies a radially increasing mass
distribution.  Moreover, Table \ref{tab:modfits} shows that envelope
masses typically exceed disk masses in best-fit models.

Table \ref{tab:massdist} lists the total (disk+envelope) masses within
100 AU for each object and model.  This scale corresponds roughly with
the resolution of our $\lambda$1.3 mm observations.  Constraints on
the mass distribution on still more compact scales are 
uncertain.  The mass found in the inner 100 AU has a median value of
0.007 M$_{\odot}$ for our sample (with large variations).  The
fractional contribution of this ``compact'' mass distribution to the
total mass  varies from 0.03 to 0.44, with a median value of 0.18.

\begin{figure}
\plottwo{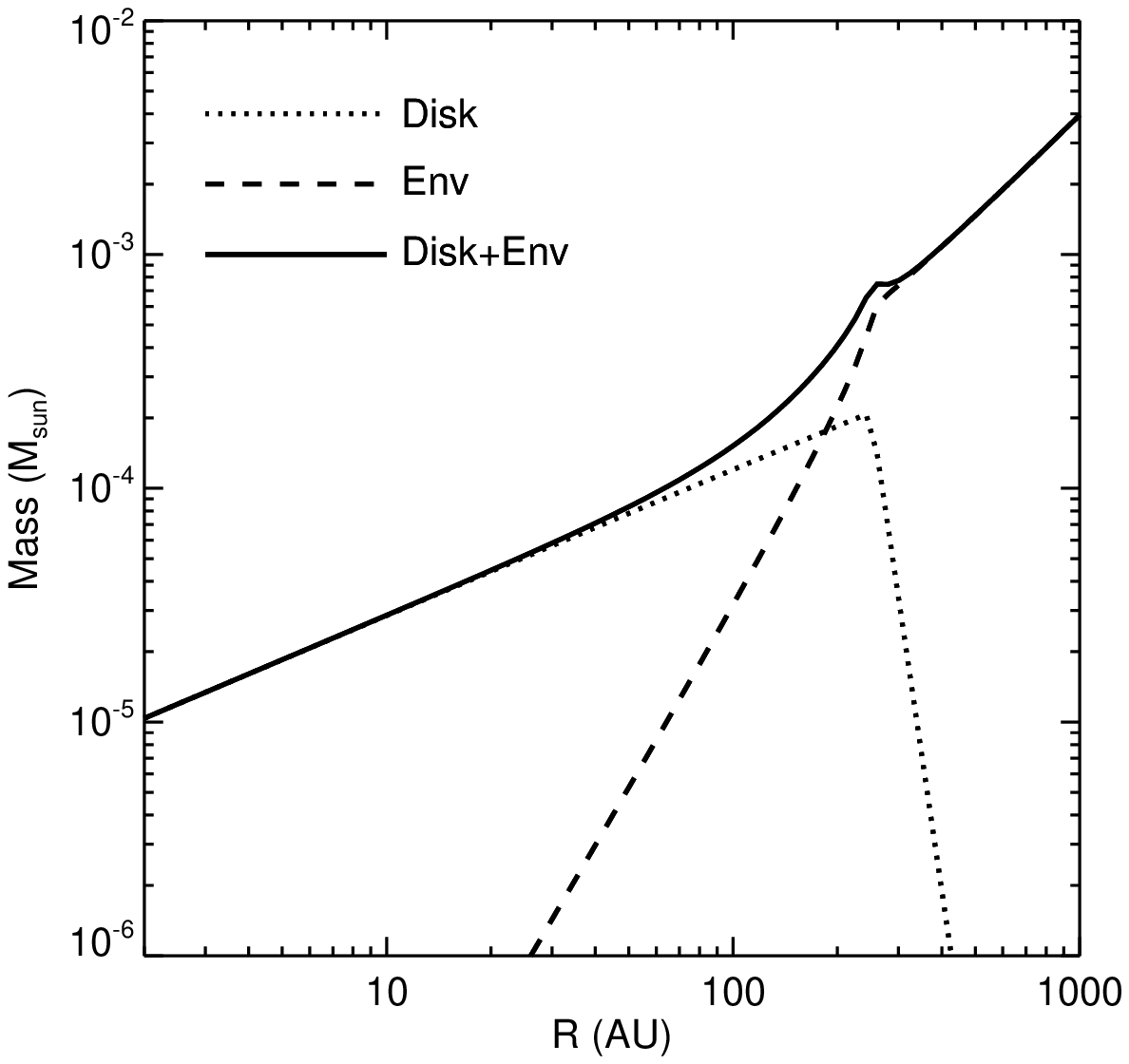}{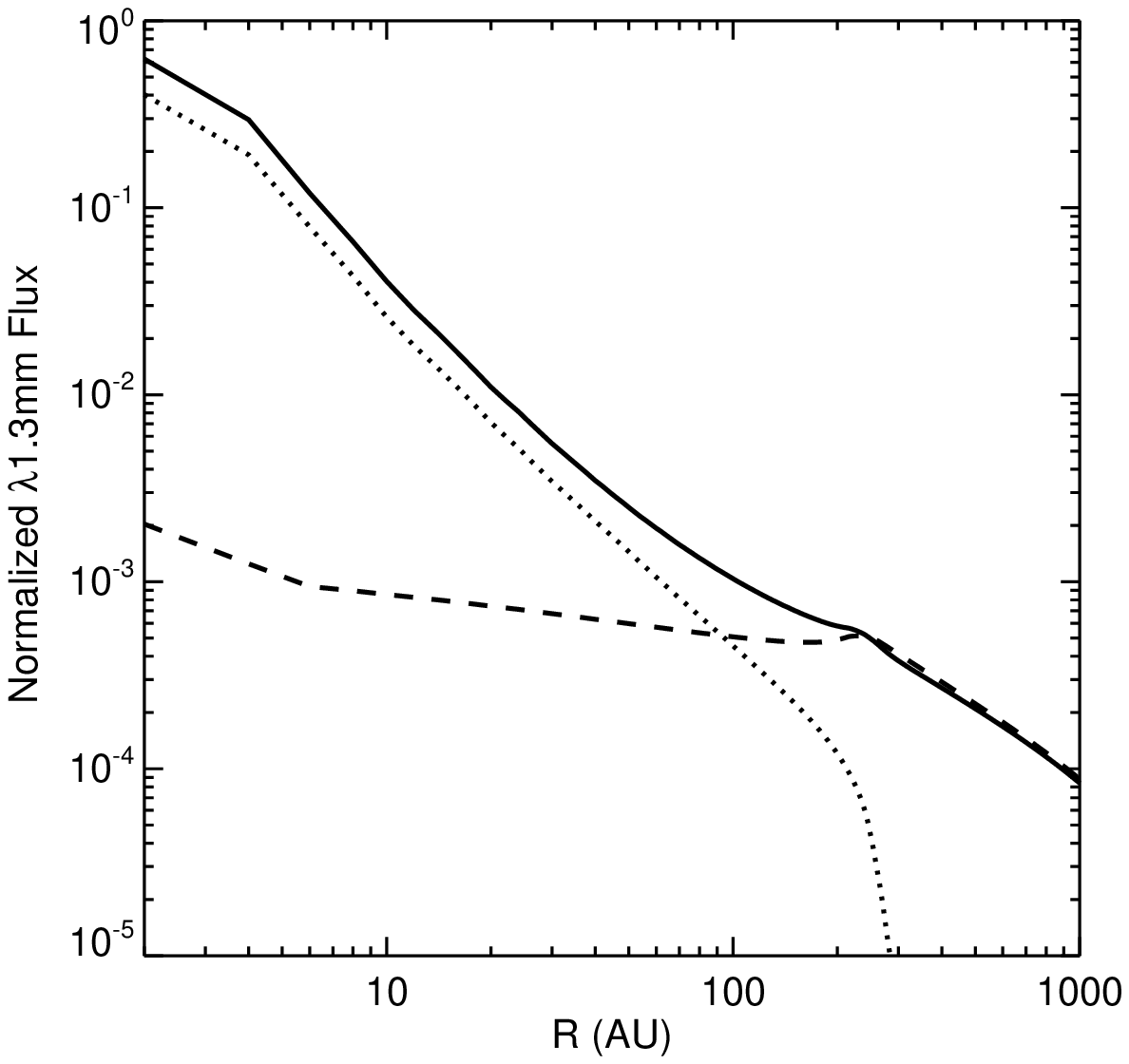}
\caption{Radial mass distribution ({\it left}) and azimuthally averaged
$\lambda$1.3 mm flux distribution ({\it right}) for the disk+envelope
model listed in the first row (model ``A'' for IRAS 04016+2610) 
of Table \ref{tab:modfits}.  The disk
component contributes most of the mass within 100 AU, although this
represents only a small percentage of the total mass.  However the
disk component dominates the observed flux at
$\lambda$1.3 mm, at least on the relatively compact scales ($\la 1000$ AU)
probed by our CARMA observations.
Note that the component fluxes were determined from
models where the disk or envelope mass was set to zero, while the flux
of the disk+envelope model is determined from radiative transfer
modeling for the combined density distribution.
\label{fig:fprofile}}
\end{figure}

Even though the mass within 100 AU is often a small fraction of the
total mass, it produces a relatively large fraction of the millimeter
flux (Figure \ref{fig:fprofile}).  The circumstellar material is
considerably hotter at small stellocentric radii.  At long wavelengths
where emission strength is approximately proportional to temperature
and circumstellar mass, even a modest temperature gradient can lead to
higher millimeter fluxes at smaller radii.  Thus our modeling can
constrain disk mass and geometry (as seen in Table \ref{tab:modfits})
even if the disk contributes modestly to the total mass.

\subsection{Comments on Individual Sources}

\subsubsection{IRAS 04016+2610} 
\begin{figure}
\plotone{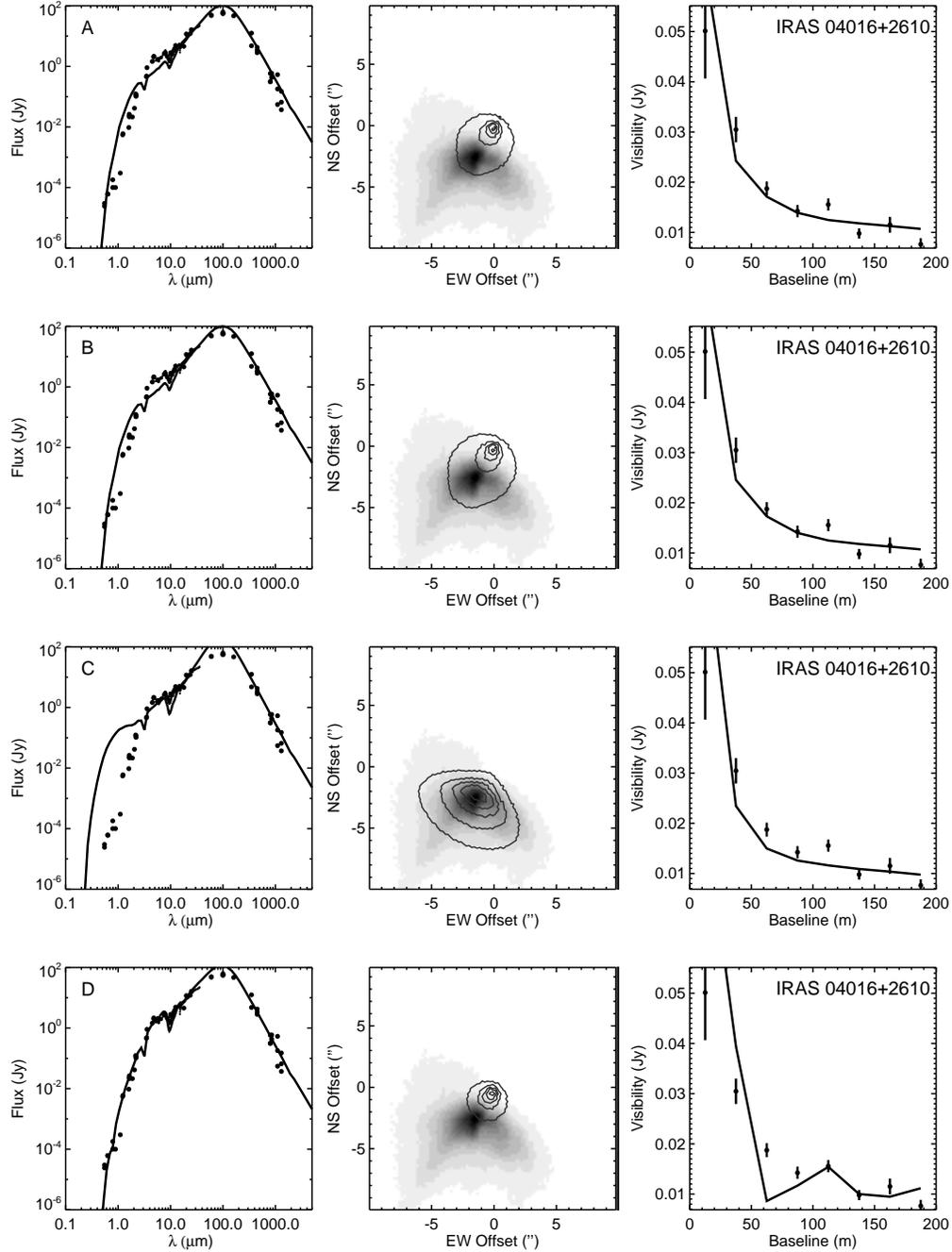}
\caption{Observed and synthetic data for IRAS 04016+2610.  We plot:
  broadband SEDs, including Spitzer/IRS spectra ({\it left panels});
  I-band scattered light images ({\it middle panels}); and
  azimuthally-averaged $\lambda$1.3mm visibilities ({\it right
    panels}).  Models are represented by solid curves in the left and
  right panels, and by contours in the middle panels.
  The different rows show best-fit models for different
  relative weightings of the datasets.  The plotted models follow the
  same ordering (top-to-bottom) as the entries in Table
  \ref{tab:modfits}, where data weights and  parameters of best-fit
  models are listed.
\label{fig:i04016}}
\end{figure}

\begin{figure}
\plotone{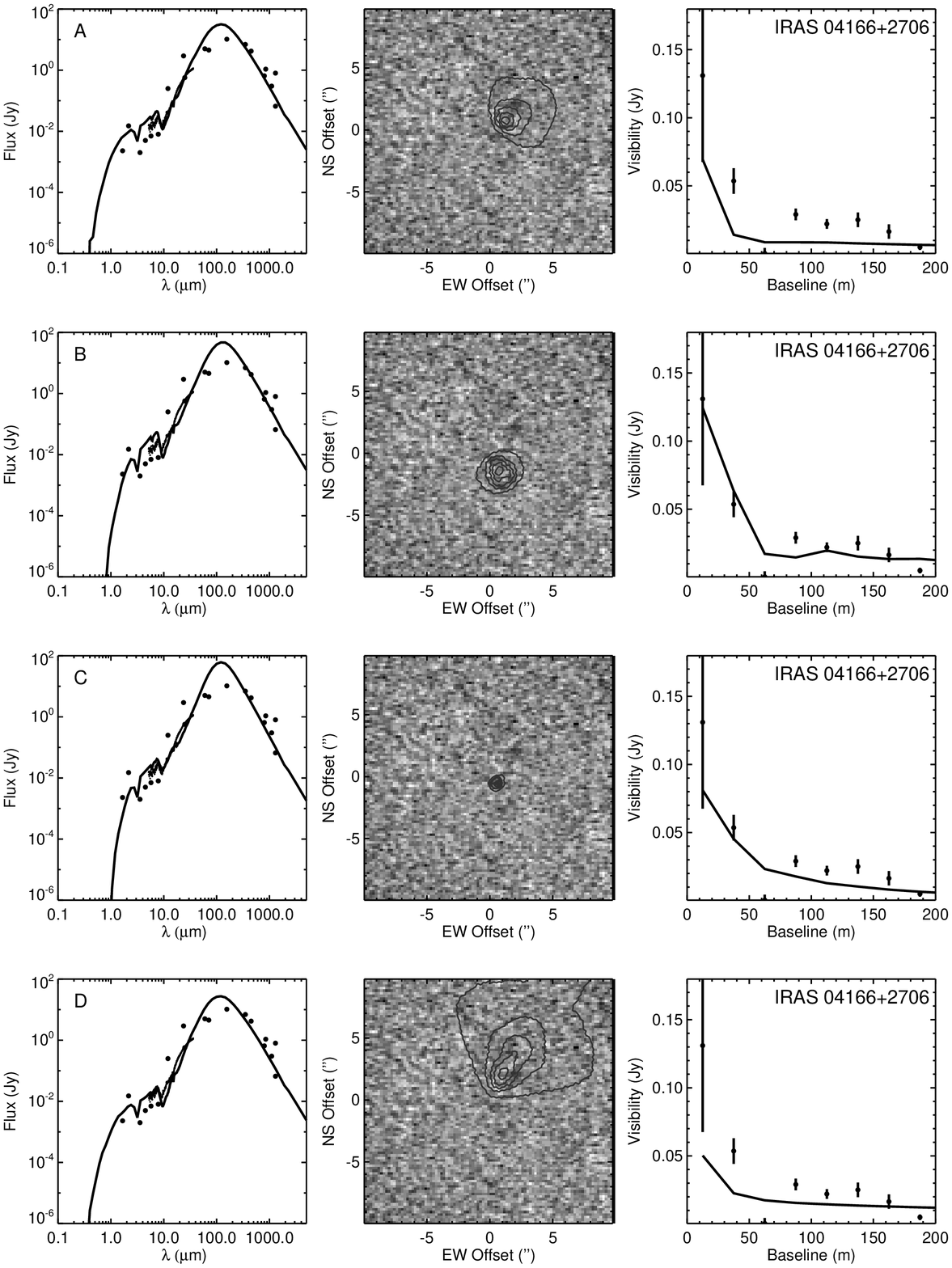}
\caption{Observed and synthetic data for IRAS 04166+2706.  See Figure
  \ref{fig:i04016} for a description of the various panels.
\label{fig:i04166}}
\end{figure}

\begin{figure}
\plotone{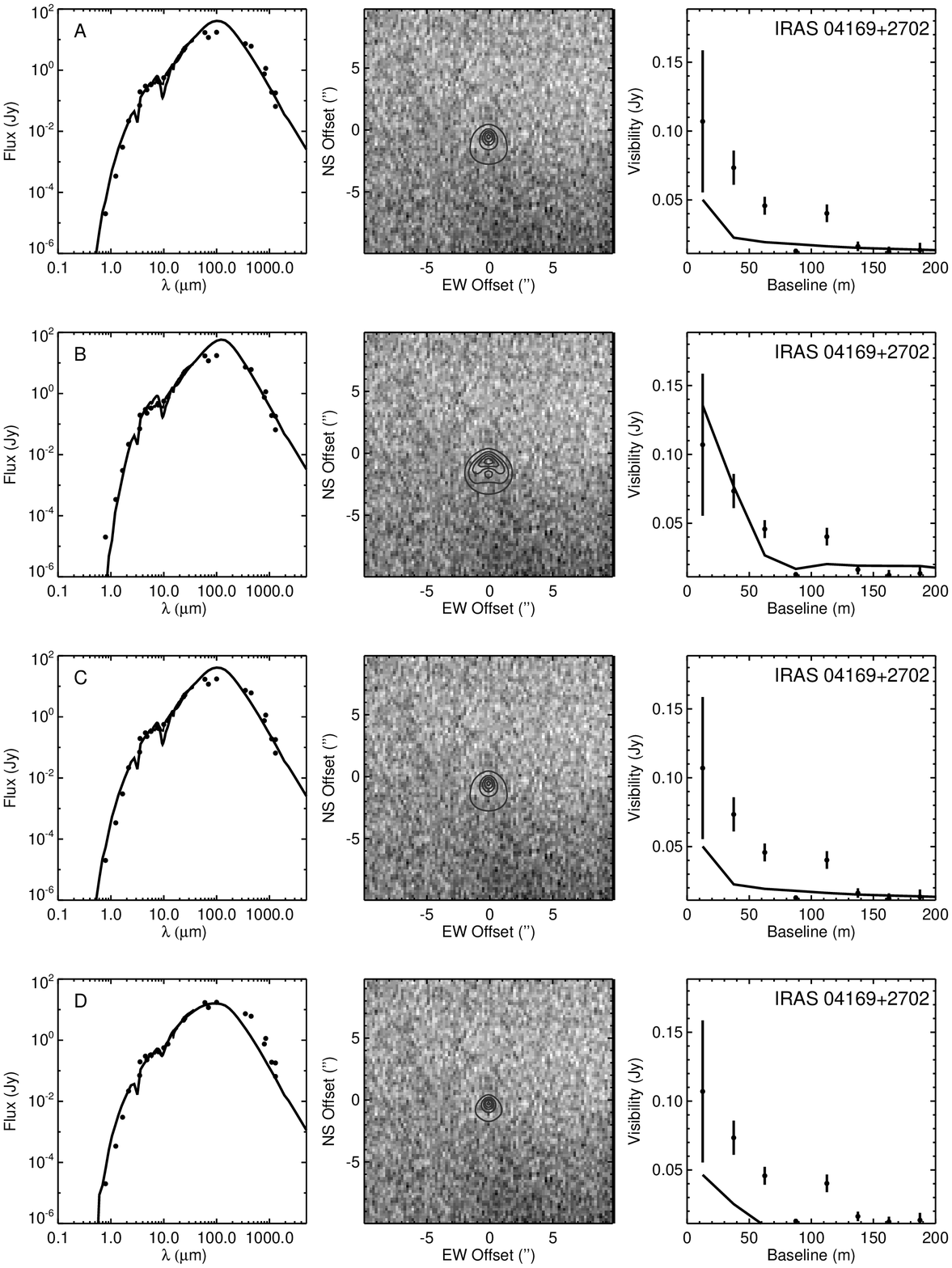}
\caption{Observed and synthetic data for IRAS 04169+2702.  
See Figure
  \ref{fig:i04016} for a description of the various panels.
\label{fig:i04169}}
\end{figure}

\begin{figure}
\plotone{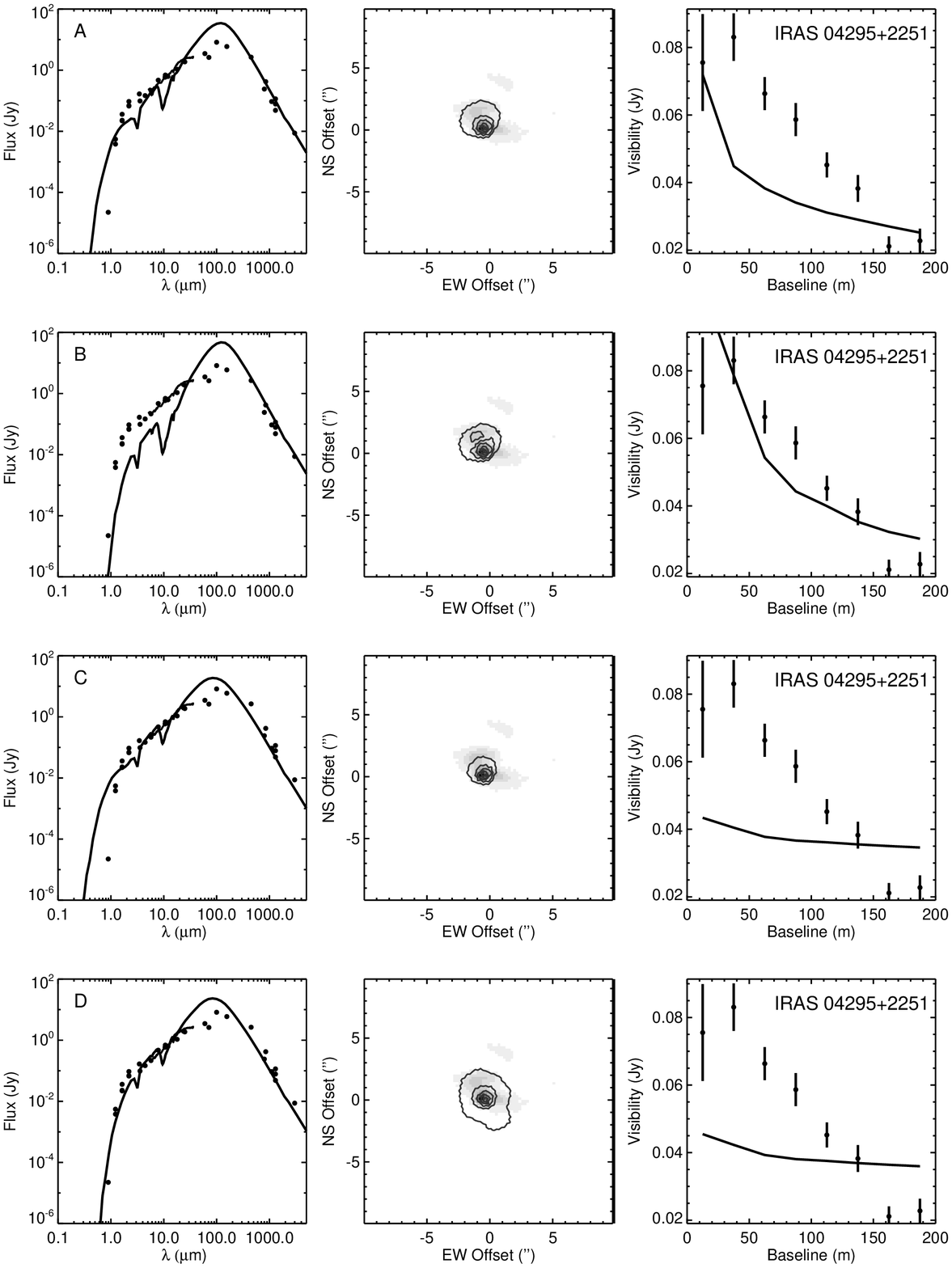}
\caption{Observed and synthetic data for IRAS 04295+2251.  See Figure
  \ref{fig:i04016} for a description of the various panels. 
\label{fig:i04295}}
\end{figure}

\begin{figure}
\plotone{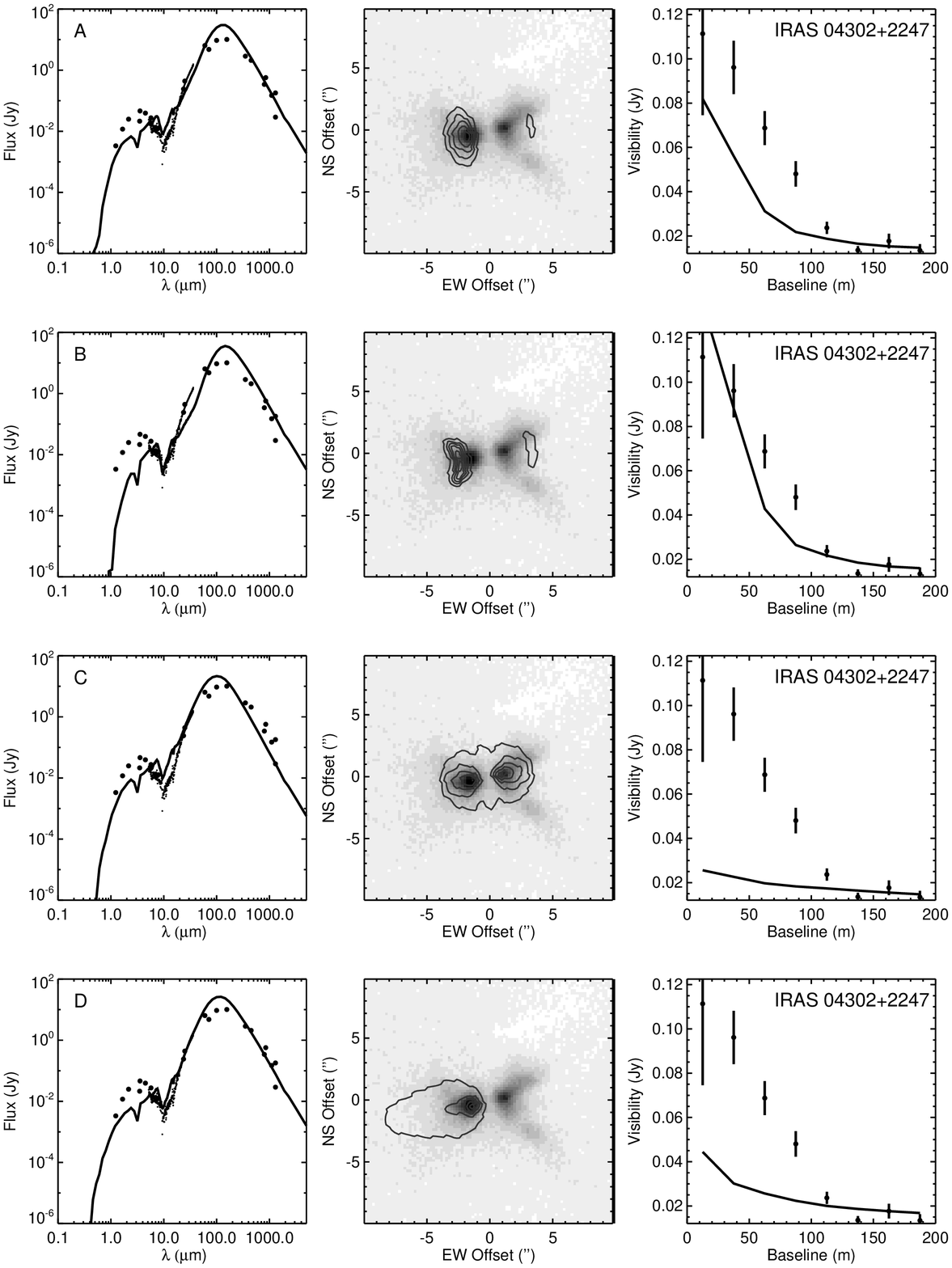}
\caption{Observed and synthetic data for IRAS 04302+2247.  See Figure
  \ref{fig:i04016} for a description of the various panels. 
\label{fig:i04302}}
\end{figure}

\begin{figure}
\plotone{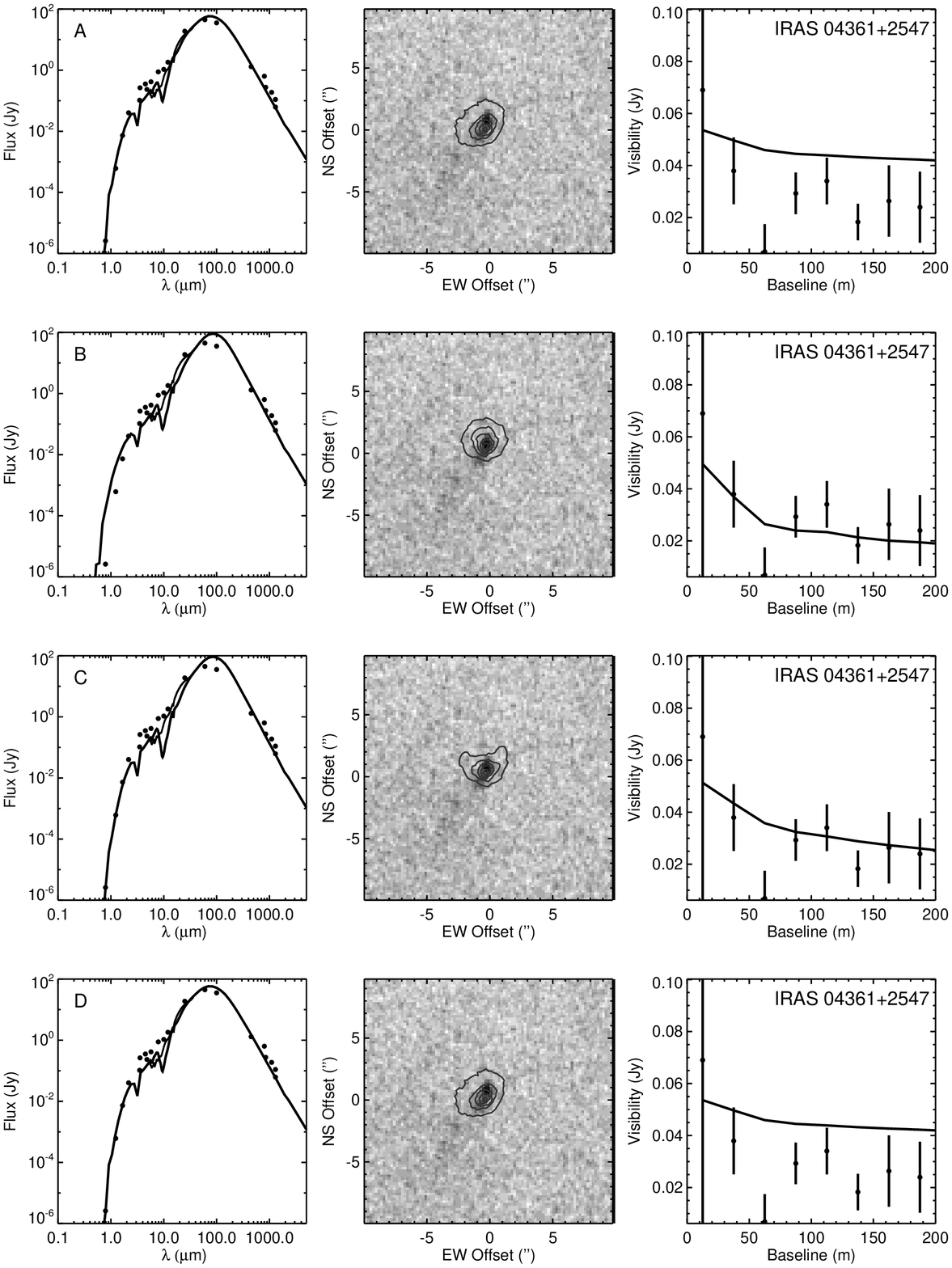}
\caption{Observed and synthetic data for IRAS 04361+2547.  See Figure
  \ref{fig:i04016} for a description of the various panels. 
\label{fig:i04361}}
\end{figure}

\begin{figure}
\plotone{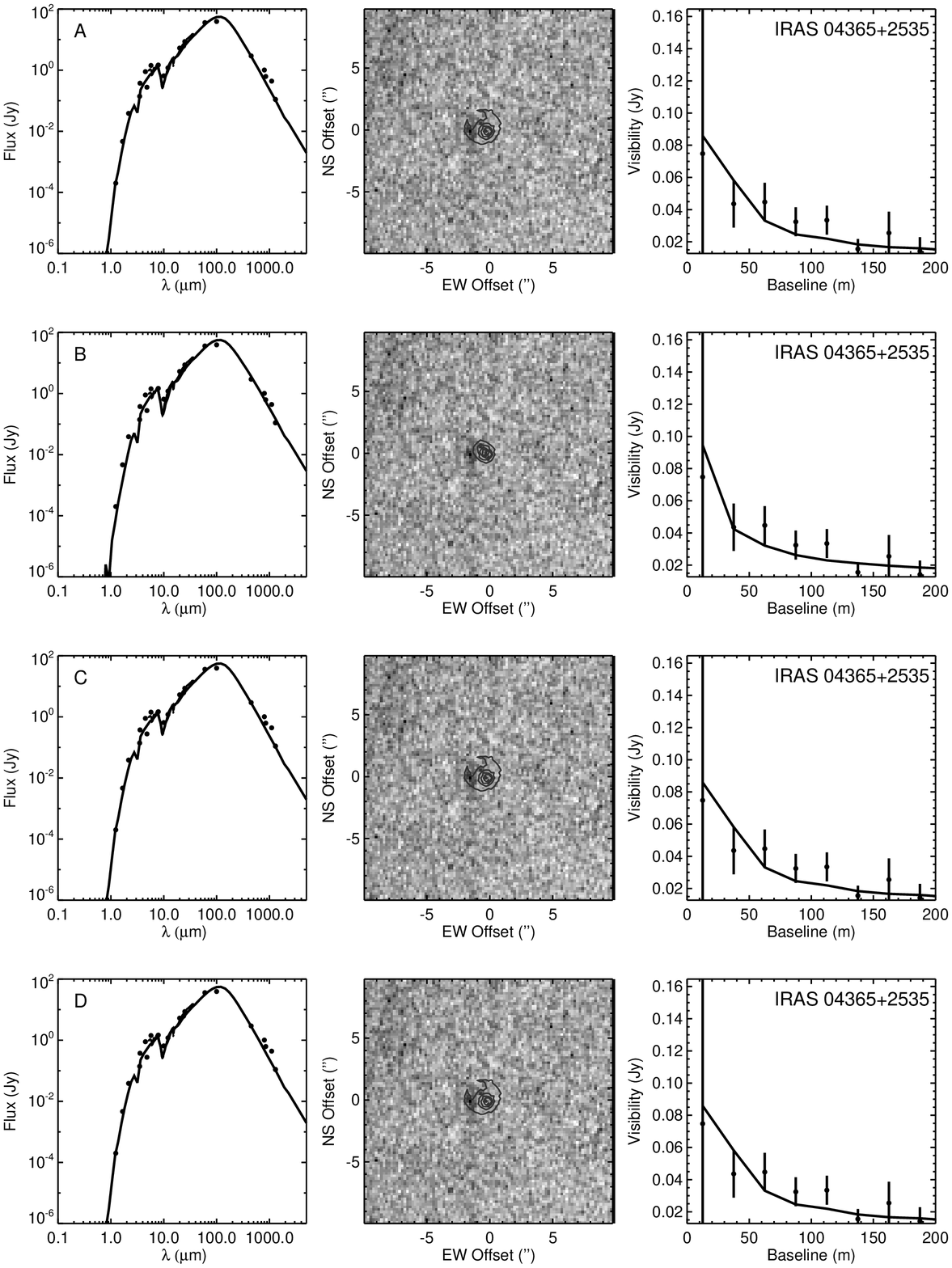}
\caption{Observed and synthetic data for IRAS 04365+2535.  See Figure
  \ref{fig:i04016} for a description of the various panels. 
\label{fig:i04365}}
\end{figure}

Models of this source generally fit all of the multi-wavelength data
well.  However, as seen in Figure \ref{fig:i04016}, fitting the
positional offset of the millimeter and I-band emission well leads to
a somewhat worse fit to the SED data (third row of Figure
\ref{fig:i04016} and Table \ref{tab:modfits}).  If we allow our
modeled scattered light image to have a smaller offset than seen in
the data, we can easily reproduce other aspects of the data.
Furthermore, the offset in the model image is affected strongly by the
outflow cavity parameter $\zeta$ (Figure \ref{fig:zeta}), and so it
seems that tweaking this parameter could improve the overall fits.

Our modeling of this source indicates a disk mass of 0.005
M$_{\odot}$, a disk radius of 250--450 AU, an envelope mass of
0.05--0.10 M$_{\odot}$, a luminosity of $\ga 3$ $L_{\odot}$, an
inclination of $\sim 35$--65$^{\circ}$, and a PA of $\sim
60^{\circ}$.  The disk scale height, outer radius, and envelope
outflow cavity properties are less well constrained.

The envelope appears to be relatively massive around this source, with
$\sim 10$ times more mass in the envelope than in the disk.  The ratio
of compactly distributed (within 100 AU) to the total mass is $\sim
0.03$ (Table \ref{tab:massdist}).  The large ratio of envelope to disk
mass suggests a younger, less evolved object.  On the other hand, the
values of $R_{\rm disk} = R_{\rm c}$ are $\ge 250$ AU, comparable to
disk sizes in T Tauri stars \citep[e.g.,][]{DUTREY+96}.  Thus, the
disk may be fully formed
in this object, despite continuing infall from an extended envelope.

Earlier modeling of millimeter wavelength continuum and line data 
suggested a higher disk mass ($\sim 0.02$
M$_{\odot}$) with a more highly inclined (60--90$^{\circ}$)  viewing
angle \citep{HOGERHEIJDE01}.  However more recent millimeter data has
been interpreted in the context of a model including disk and envelope
masses close to those derived here \citep{JORGENSEN+09}.  
Analysis of scattered light images also produces geometric parameters
consistent with those derived here \citep[e.g.,][]{GRAMAJO+07}.
This well-studied object has even been subjected to previous modeling
of combined multi-wavelength data
\citep{EISNER+05b,BRINCH+07a,BRINCH+07b}; again results of this
modeling are consistent with those presented here.
We note, finally, that the luminosity
inferred in our modeling is compatible with the effective temperature
of this source compared to other objects in our sample
\citep{DOPPMANN+05,PRATO+09,CG10}.

\subsubsection{IRAS 04166+2706}
For this object our models are able to reproduce well the observed SED
and millimeter visibilities.  Since no scattered light is detected
from this source, the model scattered light image morphologies are not
terribly important.  However models with low scattered light fluxes
are more consistent with the data.  The models shown in the second and
third rows of Figures \ref{fig:i04166} (and seen in the second and
third entries for IRAS 04166+2706 in Table \ref{tab:modfits}) have
smaller optical/near-IR fluxes.  These models still reproduce well the SED and
millimeter visibilities.

Inferred disk masses for this object are $\sim 0.01$ M$_{\odot}$.
While a heavy weighting of the scattered light imaging data produces a
disk mass of 0, we note that this model has a much smaller value of
$R_{\rm c}$ than others, which means that it still has a significant
amount of compactly distributed mass.  Indeed, the mass within 100 AU
is 0.003--0.005 M$_{\odot}$ (Table \ref{tab:massdist}) 
for all model fits to this object listed in Table \ref{tab:modfits}.

The envelope mass is $\sim 0.10$ M$_{\odot}$, and the ratio of mass
within 100 AU to the total is $\sim 0.05$.  The large ratio of
extended mass to compact mass suggests a young age.
While the large inferred disk radius ($\sim 450$ AU) suggests a
possible older age, we note that a similar fit can be achieved with no
disk at all, but with an envelope including a flattened density
distribution within $R_{\rm c} = 30$ AU.

The luminosity for this object is inferred to be $\sim 1$--3
L$_{\odot}$.  The inclination and position angle are 50-65$^{\circ}$,
and 120--240$^{\circ}$, respectively.  The large range in allowed
inclination and position angle (compared with some other objects in
our sample) is due to the lack of scattered light emission, which
provides strong constraints on geometry.  The disk scale height and
envelope outer radius are not well constrained.

The results presented here are generally compatible with previous
models of spectral energy distributions
\citep[e.g.,][]{KCH93,FURLAN+08}.  However, our study is the first to
model multi-wavelength images and the SED simultaneously.

\subsubsection{ IRAS 04169+2702}
Our models for this source provide a reasonable fit to all of the
data, although we deem the fits to be worse when we weight the
scattered light imaging or SED data heavily.  In these cases, models
tend to under-predict the observed millimeter visibilities.  However,
as seen in the second row of Figure \ref{fig:i04169}, additional
weighting of the millimeter data leads to a model that fits all
datasets well.  

The scattered light images show no detectable emission, and the weak
I-band flux in our models in compatible with this observation.  The
scattered light morphology in our models is also consistent with weak
scattered light emission seen in previous work
\citep[e.g.,][]{KENYON+93}.

The best-fit models, listed in Table \ref{tab:modfits}, yield a disk
mass of $\sim 0.01$ M$_{\odot}$, a disk radius of $\sim 250$--450 AU,
and an envelope mass of $\sim 0.1$ M$_{\odot}$.  The mass within 100
AU is $\sim 0.004$ M$_{\odot}$, which accounts for $\sim 0.04$ of the
total mass (Table \ref{tab:massdist}).  The inferred luminosity is
$\sim 1$ L$_{\odot}$, and the inclination and position angle are
30--35$^{\circ}$ and 90$^{\circ}$, respectively.  As for other objects
in our sample, the disk scale height, envelope outer radius, and
outflow cavity filling factor, are not well-constrained.

\subsubsection{IRAS 04287+1801 (L1551 IRS 5)}
\begin{figure}
\plotone{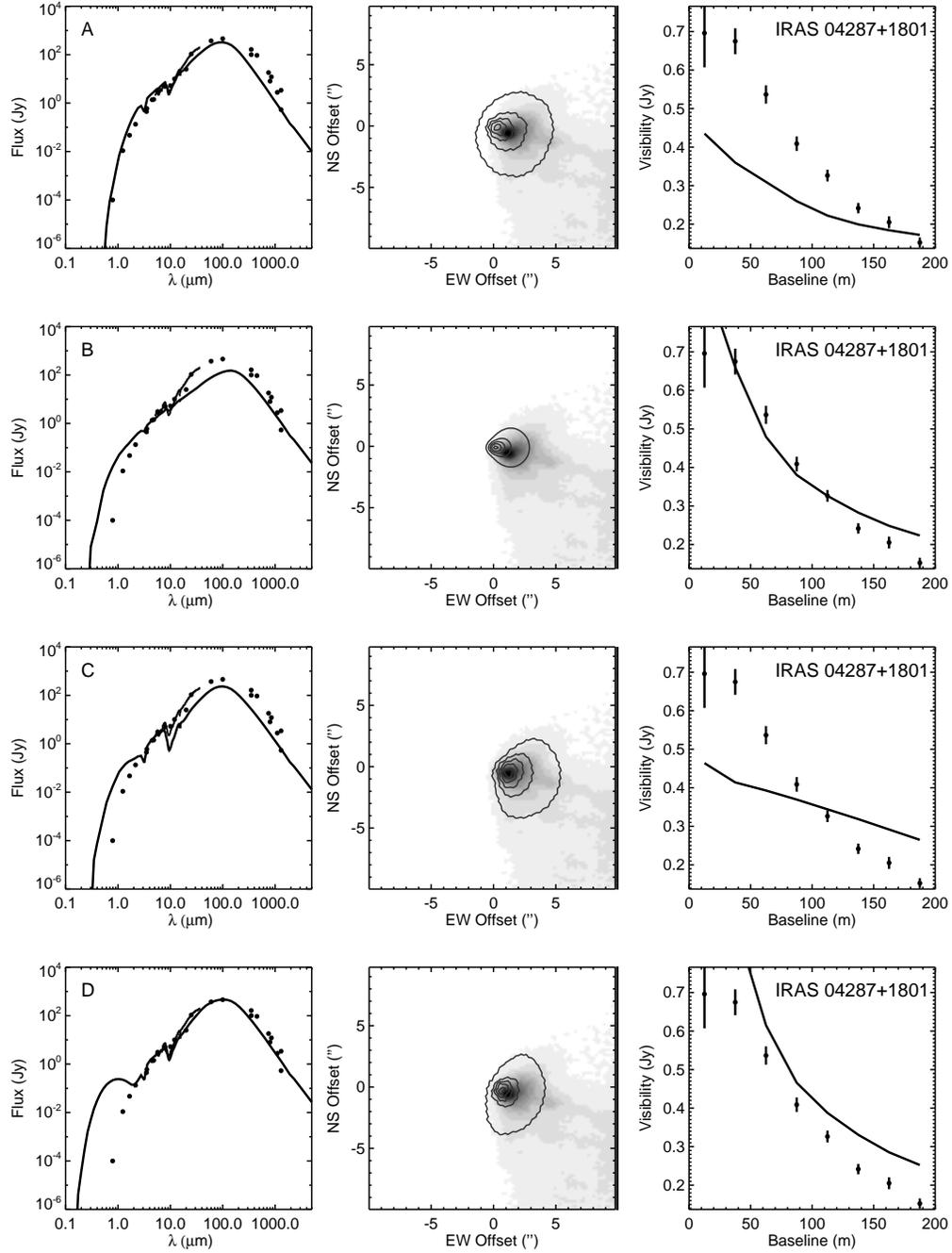}
\caption{Observed and synthetic data for IRAS 04287+1801.  
See Figure
  \ref{fig:i04016} for a description of the various panels. 
\label{fig:i04287}}
\end{figure}

As seen in Figure \ref{fig:i04287}, we were not able to provide a
perfect fit to our entire multi-wavelength dataset.  We can fit the SED
and scattered light image well together, but that model
under-predicts the millimeter visibilities  (top row of Figure
\ref{fig:i04287}).  If we fit the scattered
light image and millimeter visibilities well, the SED fit quality is
degraded (second row of Figure \ref{fig:i04287}).   We can fit all of
the data reasonably well if we allow the model to over-predict the
optical/near-IR fluxes (bottom row of Figure \ref{fig:i04287}).  Some
additional extinction, for example from a  different shape of the
outflow cavity (Figure \ref{fig:zeta}), may provide an even better fit.

The difficulty in achieving a high-quality fit to this object is due, in
part, to the sparse
sampling of parameter values in our modeling.  Best-fit models often
choose the highest sampled values of $M_{\rm disk}$, $M_{\rm env}$,
and $L_{\ast}$.  In particular, a value of $L_{\ast} > 10 L_{\odot}$ 
would produce a higher peak SED and larger overall millimeter
visibilities (Figure \ref{fig:lum}).  Thus, a luminosity larger than
10 $L_{\odot}$ would probably result in a better fit to the
data.

IRAS 04287+1801 is also known to be a multiple system, which may
complicate fitting of our simple models to the data.
This object is clearly resolved as a $\sim 0\rlap{.}''3$ 
binary \citep{RODRIGUEZ+98}, 
and even shows evidence for a third component $<0\rlap{.}''1$ away
from the northern component in high resolution 
observations \citep{LT06}.  All of these components would be within
the resolution of our CARMA observations. 

Previous authors have modeled SED data, scattered light imaging, and
even millimeter imaging data for this source.
\citet{OSORIO+03} attempted to
fit the SED and low-resolution sub-millimeter imaging with a binary
protostar model.  Their model specifies a circumbinary disk mass of
0.4 M$_{\odot}$, a disk radius of 300 AU, an envelope mass of $\sim 1$
M$_{\odot}$, $L_{\ast} = 25$ L$_{\odot}$, and $i$=50$^{\circ}$.
Subsequent modeling of the SED \citep{FURLAN+08} or scattered light
imaging \citep{GRAMAJO+07} finds similar results, but with smaller
envelope masses and disk/centrifugal radii. 

Inferred parameter values from our modeling
are: $M_{\rm disk} = 0.1$--0.5 M$_{\odot}$, $R_{\rm disk} = 100$--450
AU, $M_{\rm env} = 0.05$--0.1 M$_{\odot}$, $L_{\ast} \sim 10$
L$_{\odot}$, $i=40$--50$^{\circ}$, and PA=160--180$^{\circ}$ (Table
\ref{tab:modfits}).  These
are generally consistent with previous results, with the exception of
a lower $L_{\ast}$.  It seems that we recover most of the details of
the circumstellar mass distribution, even for this complicated
multiple system, using relatively simple disk+envelope models. 

\subsubsection{IRAS 04295+2251}
Best-fit models provide a reasonable fit to the combined dataset for
IRAS 04295+2251, although better fits to individual datasets are
possible at the expense of degrading the fits to other data (Figure
\ref{fig:i04295}).  For example, the best-fit model to the millimeter
visibility data does not fit the SED as well as other models.  This
may reflect the sparse sampling in our model grid to some extent.
However, the fact that none of our models provide an excellent fit to
the SED suggests that certain assumed parameters
 (e.g., dust opacities or outflow cavity shape) need to be adjusted to
 improve the overall fits.

Our modeling implies a disk mass of 0.01 M$_{\odot}$ and a relatively
compact disk radius of 30--100 AU.  The envelope mass is inferred to
be 0.005-0.05 M$_{\odot}$, the envelope radius is 500--1000 AU,
$L_{\ast} = 1 L_{\odot}$, and
the inclination and position angle are 45--55$^{\circ}$ and
300$^{\circ}$, respectively (Table \ref{tab:modfits}).  This source
has a high fraction of compact-to-extended mass, with $\sim 44\%$ of
the matter distributed within 100 AU (Table \ref{tab:massdist}).

While most of the inferred parameter values are similar to those found
in previous modeling of multi-wavelength data \citep{EISNER+05b}, the
disk mass presented here is substantially lower.  We find 0.01
M$_{\odot}$, while \citet{EISNER+05b} found 1 M$_{\odot}$.  However,
\citet{EISNER+05b} assumed the disk mass was distributed over radii
out to 500 AU, while the modeling presented here places the disk mass
within 30--100 AU.  Given the higher resolution of the data presented
here, it is not surprising that the disk mass is now constrained to
smaller radii.  Since the matter at smaller radii produces most of the
emission (Figure \ref{fig:fprofile}), a smaller disk requires less
total mass to reproduce the observations.

\subsubsection{IRAS 04302+2247} 
We are unable to find a single model that reproduces the combined
multi-wavelength dataset for this source.  We can fit well the SED and
millimeter data together, or the SED and scattered light data
together, but not the millimeter and scattered light data together
(Figure \ref{fig:i04302}). However, the best-fit models for different
data weightings produce fairly consistent values for some parameters.
Disk mass is 0.005--0.01 M$_{\odot}$, and the disk radius is 100--250
AU.  The envelope mass is not well constrained, spanning values
between 0.005 and 0.1 M$_{\odot}$, while the envelope outer radius
appears to be $\sim 500$ AU.  The inclination is 70--89$^{\circ}$, and
the PA is 10$^{\circ}$.  The mass within 100 AU is found to be $\sim
0.006$ M$_{\odot}$, accounting for $\sim 20\%$ of the total mass
(Table \ref{tab:massdist}).

Previous modeling of millimeter and scattered light imaging data 
argued for different dust grain properties in the disk and envelope
components \citep{WPS03}.  We have not included such effects in our
modeling, which may explain why we have difficulty fitting the
millimeter visiblities and the scattered light images simultaneously.
With a detailed model for this single object, \citet{WPS03}
derived $M_{\rm disk } =0.07$ AU, $R_{\rm disk} = 300$ AU,
$h_{\rm 1 \: AU} = 0.15$ AU, $M_{\rm env} = 0.05$ M$_{\odot}$, and
$R_{\rm out} = 450$ AU.  For the most part, our modeling (Table
\ref{tab:modfits}) is consistent with these results.  The disk mass we
infer is substantially lower, however, perhaps due to a somewhat
smaller fitted disk radius, as well as different
assumed opacities.

\subsubsection{IRAS 04361+2547}
We are able to fit the data for this source quite well, especially
when the imaging data is given some additional weight compared to the
SED data (Figure \ref{fig:i04361}).  The disk mass is inferred to be
0.005--0.01 M$_{\odot}$, and the envelope mass is comparable, $\sim
0.01$ M$_{\odot}$.  The disk radius is not well constrained, while the
outer radius of the envelope is inferred to be 500 AU.  The total
luminosity is 3 $L_{\odot}$, the inclination is $\sim 45^{\circ}$, and
the position angle is $\sim 280^{\circ}$.  This source exhibits a
relatively high fraction of compact-to-extended mass content, with
$\sim 40\%$ of the mass distributed within 100 AU (Table
\ref{tab:massdist}).

The disk mass we derive is consistent with \citet[][who found $M_{\rm
  disk} = 0.009$ M$_{\odot}$]{JORGENSEN+09}, but
the envelope mass we infer
here is somewhat lower.  This is probably a consequence of the fact that
our millimeter observations are not sensitive to emission more
extended than $\sim 1000$ AU, while \citet{JORGENSEN+09} used
single-dish observations to detect more extended emission.
We predict a somewhat lower inclination than
found in previous scattered light imaging
\citep{KENYON+93,GRAMAJO+07}; however yet lower inclination is
inferred in previous SED modeling \citep{KCH93}.  Thus our inferred
geometry is generally consistent with previous work.

\subsubsection{IRAS 04365+2535}
Our models for IRAS 04365+2535 fit the combined multi-wavelength
dataset well, regardless of data weighting factors (Figure \ref{fig:i04365}).
Best-fit models imply $M_{\rm disk} = 0.005$ M$_{\odot}$, $R_{\rm
  disk} = 100$ AU, $M_{\rm env} = 0.05$--0.1 M$_{\odot}$, $R_{\rm out}
= 500$--1000 AU, $L_{\ast} = 1$ $L_{\odot}$, inclination$=25^{\circ}$,
and position angle$=310^{\circ}$.  The mass enclosed within 100 AU is
$\sim 0.007$ M$_{\odot}$, which comprises $\sim 11\%$ of the total
mass.

Previous analysis of millimeter visibility data determined 
a disk mass of 0.03 M$_{\odot}$ and an envelope mass of 0.12
M$_{\odot}$  \citep{JORGENSEN+09}.   
The disk mass in particular is higher than we infer
in our analysis.  This discrepancy is probably due to the lower
angular resolution in the previous observations:  $\sim 2\rlap{.}''5 =
350$ AU.  In contrast, we infer a disk size of
100 AU.  With a larger beam, the earlier observations could have
easily inferred a larger mass.  The results of our modeling also
appear consistent with previous modeling of the SED or scattered light
imaging \citep{KCH93,KENYON+93,GRAMAJO+07}, although our inferred
inclination is somewhat lower than previous estimates.

\section{Discussion}
The disk mass is a crucial parameter in understanding both star and
planet formation.  High disk masses, on the order of a tenth of the
stellar mass, would imply that gravitational instability may be important.
Such instability can lead to rapid accretion onto the central
protostar, and may explain why accretion rates onto the star appear
low compared to estimated time-averaged accretion from the envelope
onto the disk.  Episodes of gravitationally-enhanced accretion can lead
to larger time-averaged rates but low typical instantaneous accretion
rates \citep[e.g.,][]{WH04,EISNER+05b}.

Disk mass is also a critical initial condition for planet formation.
To form gas giants like Jupiter and Saturn requires $>0.01$
M$_{\odot}$ of matter \citep{WEID+77}.  While measured disk masses
around T Tauri stars are smaller 
\citep[e.g.,][]{AW05,EISNER+08}, this may reflect the agglomeration of
small particles into larger bodies rather than a mass deficit.  Data
for the
less evolved protostars in our sample can provide masses in systems
where dust grain agglomeration is presumably less advanced.

We begin by examining the distribution of disk masses for our sample,
drawing values from Table \ref{tab:modfits}.  Disk masses range from
0.005 to 0.5 M$_{\odot}$ across the sample, and the median disk mass is
0.008 M$_{\odot}$.   The majority ($\sim 60\%$) of objects have disk
masses $\ge 0.01$ M$_{\odot}$, and $\sim 10\%$ have $M_{\rm disk} \ge
  0.1$ M$_{\odot}$.
For comparison, across $>150$ (more evolved)
Class II disks in Taurus, \citet{AW05} find a median mass of 0.005
M$_{\odot}$.  In this sample of Class II objects, $\sim 20\%$ of disks
are more massive than 0.01 M$_{\odot}$ and only $\sim 2\%$ are more
massive than 0.1 M$_{\odot}$ \citep{AW05}.    

Our sample of Taurus Class I
objects has significantly more massive disks than the Class II
population.  Indeed, many of them may still possess sufficient mass in
small ($\sim $mm-sized) dust grains for the eventual formation of planetary
systems like our own (depending on how the mass is distributed on
scales smaller than  100 AU). The lower masses in Class II disks
likely reflect that some of the  mass of small particles available at
the Class I stage has grown into large ($>$ meter-sized) bodies within
$\sim 1$ Myr.

This argument can easily be extended to Class I sources themselves.
While Class I disks appear more massive than Class II disks---at least
as traced by small ($\la$ mm-sized) dust grains---particles in Class I
disks may already have undergone substantial processing.  That is,
planet formation may already be underway even at these ages.
Extending the analysis in this paper to younger, less evolved Class 0
objects provides a clear path to improving the estimates of the
initial mass content of protoplanetary disks.

The range of disk masses observed for our sample indicates that few 
Class I disks are at the limit of gravitational instability.  Even
IRAS 04287+1801, which has a disk mass $\ge 0.1$ M$_{\odot}$, may
remain stable against gravitational collapse because of the higher
stellar mass contained in the multiple central system.  
If, however, substantial mass is hidden in larger bodies,
gravitational instability may still be possible in these systems.

Measured envelope masses can be used to estimate the mass infall rate
from the envelope onto the disk, $\dot{M}$.  Estimates of $\dot{M}$
can then be compared to the accretion rate measured from the inner
disk onto the star \citep[e.g.,][]{WH04}.  Such comparisons typically
conclude that the infall rate from the envelope is higher than the
accretion rate onto the star, suggesting that matter piles up in the
disk and periodically accretes quickly in bursts
\citep[e.g.,][]{KH87,WH04,EISNER+05b,VB05}.  The results presented
here are generally compatible with this view.

However the envelope masses estimated here are highly uncertain.
For example, the range of $M_{\rm env}$ allowed for
IRAS 04016+2610 (Table \ref{tab:modfits}) correspond to $\dot{M}$
between $\sim 10^{-6}$ and $10^{-5}$ M$_{\odot}$
yr$^{-1}$.  Much of this uncertainty is due to degeneracies with other
parameters (primarily $R_{\rm out}$; see \S \ref{sec:parstudy}).
Additional uncertainties arise because of the limited $uv$ coverage
(i.e., range of telescope baseline separations) in our millimeter
observations.   The shortest baselines correspond to linear
resolutions of $\sim 1000$ AU at the distance to our targets.  More
extended emission is largely resolved out.  Thus, we are not sensitive
to extended emission and, as a result, our total envelope masses are
likely underestimated.

Despite uncertainties in inferred values of $M_{\rm env}$, clear
differences in the ratio of $M_{\rm disk}/M_{\rm env}$ are observed
across our sample.  The disk-to-envelope mass ratio ranges from $<0.1$
for some objects (IRAS 04016+2610) to $>1$ for others (IRAS
04287+1801).  In most cases, the disk-to-envelope mass ratio is around
0.1.  This suggests our sample is at an earlier evolutionary stage
than Class II objects, which are essentially 100\% disk.  Furthermore,
one might be tempted to assign relative ages to sample objects based
on $M_{\rm disk}/M_{\rm env}$ values.

Another method to constrain relative ages in our sample is to examine the
disk/centrifugal radii.  For collapsing, rotating envelopes, the
centrifugal radius grows with time
because of the inside-out nature of the collapse. 
For the envelope models used here, one expects the centrifugal, and
hence disk, radius to grow with time as $R(t) \propto t^3$.
If our sample objects had similar initial specific angular momenta 
(which is by no means
guaranteed), then the inferred centrifugal radii can be used to
estimate relative ages.

The disk radii inferred in our modeling have substantial
uncertainties, but do reveal trends between objects (Table
\ref{tab:modfits}).  For example, IRAS 04016+2610 seems to have a
larger disk/centrifugal radius than IRAS 04295+2251.  However, IRAS
04016+2610 has a lower disk-to-envelope mass ratio than
IRAS 04295+2251.  While a smaller disk/centrifugal radius suggests a
younger source, a larger ratio of disk-to-envelope mass suggests an
older object.  We therefore do not place much confidence in our
ability to distinguish ages amongst our sample.

To compare our inferred disk masses with expected minimum masses for
planet formation theory, we turn now to the masses listed in Table
\ref{tab:massdist}.  The Table lists circumstellar masses enclosed
within 100 AU, including both disk and inner envelope components.
Since the inner regions of the envelope have a flattened density
distribution (Equation \ref{eq:rhoenv}), they may appear disk-like,
and may contribute to the mass budget for planet formation.  

We choose 100 AU because this is approximately the smallest scale we
can resolve with our millimeter imaging data. However we would ideally
like to probe smaller scales, since planet formation theories usually
require a minimum mass within  radii or 30--50 AU
\citep[e.g.,][]{WEID+77,HAYASHI81}.  Thus, even our 100-AU-masses
(Table \ref{tab:massdist}) may over-estimate the mass available for
planet formation.  Even so, the masses listed in Table
\ref{tab:massdist} are typically lower than the 0.01--0.1 M$_{\odot}$
usually required to form planetary systems like ours.  As discussed
above, this may reflect growth of small, observable dust grains into
larger bodies.  Probing younger sources may provide a clearer picture
of how much mass is truly available to planet formation.

\section{Summary and Future Directions}
We described new millimeter imaging observations of a sample of Class
I protostars, obtained with the CARMA interferometer.  
We presented basic properties of the millimeter
emission, and a more detailed analysis of the 
millimeter data in conjunction with imaging and flux data at other
wavelengths.
We constructed models consisting of rotating, infalling envelopes and
embedded, compact disks, and generated a 
grid of synthetic multi-wavelength datasets using the
radiative transfer code RADMC.  
We fitted these models to the multi-wavelength
dataset for each target.

Our focus in this study is measuring disk and envelope masses.
We found disk masses for our sample ranging from
$\la 0.01$ to $\ga 0.1$ M$_{\odot}$.  The compact disks typically only
comprise a small percentage of the total mass ($\la 10\%$), although
this percentage is higher in a few objects.  
Envelope masses---and by extension, the disk-to-envelope mass
ratios---are usually poorly constrained by our modeling.
This is due, in large part, to the limited sensitivity of our
observations to emission more extended than $\sim
1000$ AU.  Better $uv$ coverage, for example combining extended
and compact configurations of CARMA or ALMA, will alleviate this
problem in the future.

The disk masses inferred for our sample are significantly higher than for
samples of more evolved Class II objects.  
Thus, Class I disk masses probably provide
a more accurate estimate of the initial mass budget for star and
planet formation, before particle growth and planetesimal formation
lock up mass in hard-to-see larger bodies.  However, such particle
agglomeration may already be underway by the Class I epoch, and so
measured disk masses may still be underestimates.  
Observing Class I sources at multiple millimeter and sub-millimeter
wavelengths with ALMA will help to constrain grain properties and how
these depend on location in the disk, 
providing tighter constraints on disk masses and the extent of grain
growth.

Another obvious extension of this project is to observe larger samples of
protostars over a range of evolutionary states.  In particular,
observations and modeling of Class 0 protostars can reveal if their
disk masses are larger than those inferred around Class I objects.
Comparing disk masses for samples of objects at different ages will
constrain how the mass in small dust grains evolves, and give a better
estimate of the true initial disk mass distribution.

The angular resolution
of our observations is sufficient to constrain the disk mass
distribution within $\sim 100$ AU.  In contrast, planet formation
models generally only consider the mass within 30--50 AU.  Future
work, using more extended array configurations of CARMA or ALMA, can
reach these scales and provide constraints on disk masses that are
directly comparable to critical initial conditions of planet formation
theories.

\section*{Acknowledgments}
The author gratefully acknowledges support from an Alfred P. Sloan
Research Fellowship.
He also acknowledges the efforts of Stephanie Cortes during early
stages of this work.  He is grateful to Kees Dullemond for making
his code, RADMC,  available, and to Klaus Pontoppidan for providing
dust opacities.  Elise Furlan kindly provided IRS
spectra for each source.  Discussions with Kaitlin Kratter provided
ideas for some of the analysis presented in this work.
Support for CARMA construction was derived from the states of
California, Illinois, and Maryland, the James S. McDonnell Foundation,
the Gordon and Betty Moore Foundation, the Kenneth T. and Eileen
L. Norris Foundation, the University of Chicago, the Associates of the
California Institute of Technology, and the National Science
Foundation. Ongoing CARMA development and operations are supported by
the National Science Foundation under a cooperative agreement, and by
the CARMA partner universities.


\clearpage

\clearpage

\clearpage

\epsscale{0.8}

\epsscale{1.0}

\clearpage 

\begin{figure}
\plotone{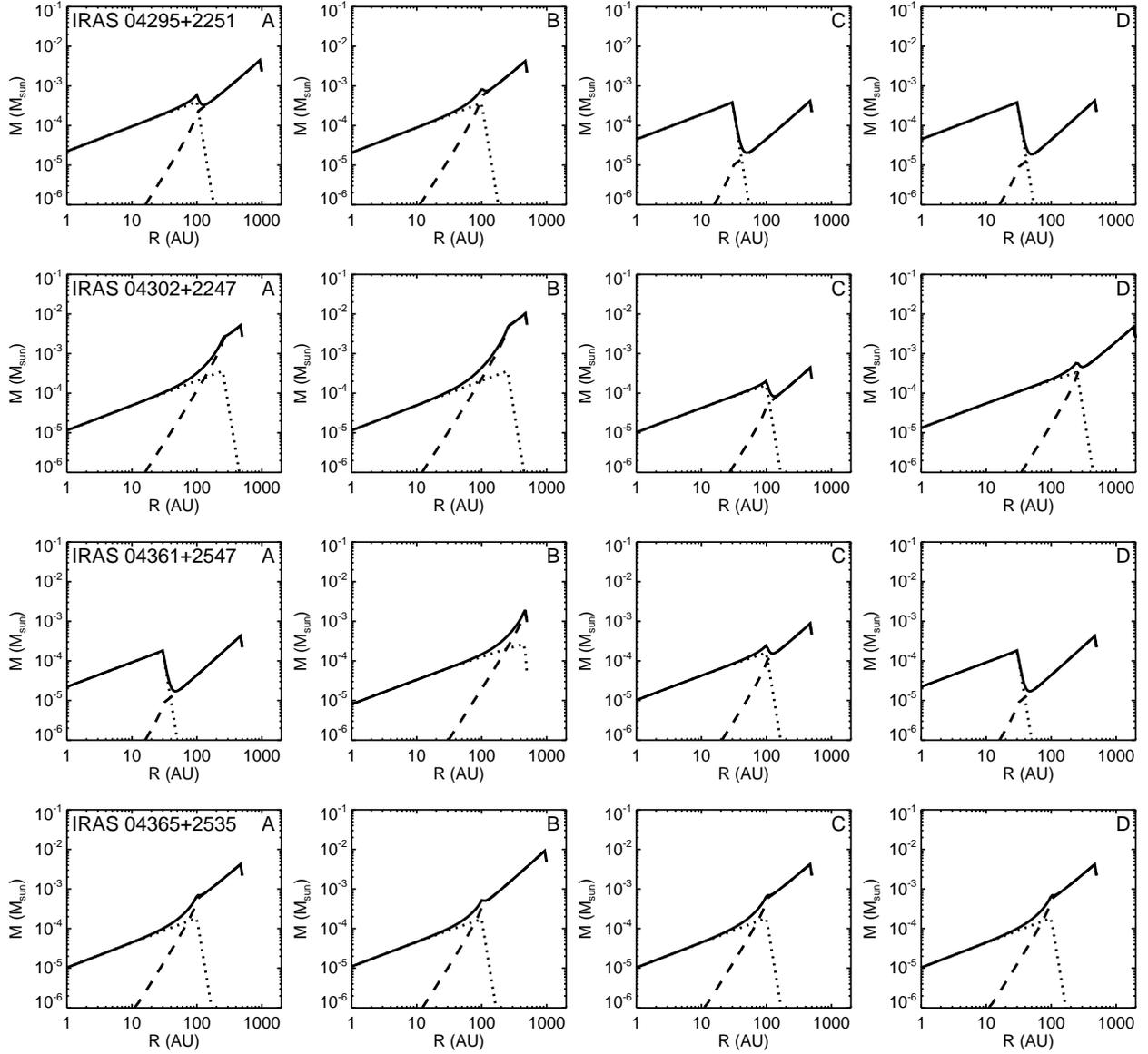}
\caption{Figure \ref{fig:mprofiles} continued.}
\end{figure}

\clearpage

\begin{deluxetable}{lcccccc}
\tabletypesize{\scriptsize}
\tablewidth{0pt}
\tablecaption{Log of Observations
\label{tab:obs}}
\tablehead{\colhead{Source} &\colhead{Observation Date}  &
  \colhead{Baselines} & \colhead{Passband Calibrators} & \colhead{Gain Calibrator} \\
 & (UT) & (m) & &}
\startdata
IRAS 04016+2610 & 2008 Oct 18, 22; 2008 Nov 17 & 18--199 & 3C84, 3C454.3 & 3C111 & \\
IRAS 04166+2706 & 2010 Oct 27 & 20-251 &  3C279, 3C273 & 3C111 \\
IRAS 04169+2702 & 2010 Oct 27 & 20-250 & 3C279, 3C273 & 3C111 \\
IRAS 04181+2654 & 2010 Oct 27 & 20-250 & 3C279, 3C273 & 3C111 \\
IRAS 04263+2426 & 2010 Oct 29 & 19-251 & 3C84, 3C111 & 3C111 \\
IRAS 04287+1801 & 2008 Oct 21, 22 & 18--197 & 3C84, 3C454.3 & 3C111 & \\
IRAS 04295+2251 & 2008 Oct 18, 22; 2008 Nov 17 & 17--198 & 3C84, 3C454.3 & 3C111 &\\
IRAS 04302+2247 & 2008 Oct 21, 22 & 18--199 & 3C84, 3C454.3 & 3C111 &\\
IRAS 04361+2547 & 2010 Oct 29 & 19-251 & 3C84, 3C111 & 3C111 \\
IRAS 04365+2535 & 2010 Oct 29 & 19-232 & 3C84, 3C111 & 3C111 \\
\enddata
\end{deluxetable}

\begin{deluxetable}{lcccccccc}
\tabletypesize{\scriptsize}
\tablewidth{0pt}
\tablecaption{Properties of 1.3-mm wavelength images
\label{tab:mmims}}
\tablehead{\colhead{Source} &\colhead{$\alpha$} & \colhead{$\delta$}
& \colhead{Peak Flux}  & \colhead{RMS} & \colhead{$\theta_{\rm maj}$}
& \colhead{$\theta_{\rm min}$} & \colhead{$\theta_{\rm PA}$} \\
 & (J2000) & (J2000) & (mJy) & (mJy) & ($''$) & ($''$) & ($^{\circ}$)}
\startdata
IRAS 04016+2610 & 04:04:43.07 & +26:18:56.3 & 17 & 1 & 1.13 &
0.85 & 54. \\
IRAS 04166+2706 & 04:19:42.50 & +27:13:36.0 & 66 & 13 & 1.03 & 0.90
& 74.0 \\
IRAS 04169+2702 & 04:19:58.48 & +27:09:57.0 & 65 & 8 & 1.02 & 0.90 & 75.9 \\
IRAS 04181+2654 & ... & ... & ... & 5 & 1.03 & 0.90 & 75.4 \\
IRAS 04263+2426N & 04:29:23.72 & +24:33:01.0 & 37 & 4 & 1.02 & 0.85
& -72.5 \\
IRAS 04263+2426S & 04:29:23.74 & +24:32:59.8 & 31 & 4 & 1.02 & 0.85
& -72.5 \\
IRAS 04287+1801 & 04:31:34.16 & +18:08:04.8 & 404 & 4 & 1.09 & 0.86
& 44.9 \\
IRAS 04295+2251 & 04:32:32.07 & +22:57:26.3 & 53 & 1 & 1.13 & 0.86 &
53.5 \\
IRAS 04302+2247 & 04:33:16.49 & +22:53:20.3 & 46 & 2 & 1.07 & 0.86 &
49.3 \\
IRAS 04361+2547 & 04:39:13.90 & +25:53:20.44 & 50 & 4 & 1.03 & 0.84
& -69.7 \\
IRAS 04365+2535 & 04:39:35.21 & +25:41:44.2 & 147 & 5 & 1.04 & 0.83
& -68.6 \\
\enddata
\tablecomments{Coordinates indicate the positions of the peak fluxes.
  The major axis, minor axis, and position angle (east-of-north) of
  the beam FWHM are listed as $\theta_{\rm maj}$, $\theta_{\rm min}$,
  and $\theta_{\rm PA}$, respectively.}
\end{deluxetable}

\begin{deluxetable}{lcccccc}
\tabletypesize{\scriptsize}
\tablewidth{0pt}
\tablecaption{Uniform Disk Model Fits to CARMA Visibility Amplitudes
\label{tab:udfits}}
\tablehead{\colhead{Source} & \colhead{Flux} &
  \colhead{$\theta_{\rm UD}$} & \colhead{$i$} & \colhead{PA} \\
 & (mJy) & ($''$) & ($^{\circ}$) & ($^{\circ}$)}
\startdata
IRAS 04016+2610 & $18 \pm 2$ & $ 2.49 \pm 0.19 $ & $   68 \pm    7 $ &
$   66 \pm    4 $ \\
IRAS 04166+2706 & $36 \pm 8$ & $ 0.75 \pm 1.00 $ & $   90 \pm   90 $ &
$  130 \pm  130 $ \\
IRAS 04169+2702 & $71 \pm 16$ & $ 1.30 \pm 0.64 $ & $   50 \pm   50 $
& $    0 \pm  180 $ \\
IRAS 04287+1801 & $536 \pm 35$ & $ 1.61 \pm 0.07 $ & $   55 \pm    9 $
& $  167 \pm    6 $ \\
IRAS 04295+2251 & $78 \pm 8$ & $ 2.05 \pm 0.12 $ & $60 \pm 7$ & $73
\pm 7$  \\
IRAS 04302+2247 &  $29 \pm 15$ & $2.76 \pm 1.26$ & $67 \pm 22$ & $33
\pm 20$ \\
IRAS 04361+2547 &  $62 \pm 28$ & $1.76 \pm 0.84$ & $60 \pm 60$ & $160
\pm 160$ \\
IRAS 04365+2535 &  $110 \pm 48$ & $2.14 \pm 0.72$ & $89 \pm 44$ & $84
\pm 25$ \\
\enddata
\tablecomments{For each source, we list the flux, angular diameter
  ($\theta_{\rm UD}$), inclination, and position angle for best-fit
  inclined uniform disk models.}
\end{deluxetable}

\begin{deluxetable}{l|ccccc|cccccccccccc}
\tabletypesize{\scriptsize}
\tablewidth{0pt}
\tablecaption{Best-fit Disk+Envelope Models
\label{tab:modfits}}
\tablehead{\colhead{Source} & & \colhead{$w_{mm}$} & \colhead{$w_{\rm
      Iband}$} & \colhead{$w_{\rm SED}$} &
  \colhead{$\chi_r^2$} &  
  \colhead{$M_{\rm disk}$} & \colhead{$R_{\rm disk}$} & \colhead{$h_{\rm 1
       AU}$} & \colhead{$M_{\rm env}$} & \colhead{$R_{\rm out}$} &
  \colhead{$f_{\rm cav}$} & \colhead{$L_{\ast}$} & \colhead{$i$} &
      \colhead{PA} \\
 &  & & & & (M$_{\odot}$) & (AU) & (AU) &  (M$_{\odot}$) & (AU) & &
 (L$_{\odot}$) & ($^{\circ}$) & ($^{\circ}$)}
\startdata
IRAS 04016+2610 & A & 1 & 1 & 1 & 5.0 & 0.005 & 250 & 0.05 & 0.10 &
2000 & 1 & 3 & 35 & 60 \\ 
IRAS 04016+2610 & B & 10 & 1 & 1 & 3.0 & 0.005 & 250 & 0.05 & 0.10 &
2000 & 1 & 3 & 40 & 60 \\  
IRAS 04016+2610 & C & 1 & 10 & 1 & 3.9 & 0.005 & 450 & 0.05 & 0.05 & 1000
& 0.2 & 10 & 65 & 60 \\ 
IRAS 04016+2610 & D & 1 & 1 & 10 & 2.8 & 0.005 & 450 & 0.15 & 0.05 & 500
& 0.2 & 3 & 35 & 60 \\ 
\hline
IRAS 04166+2706 & A & 1 & 1 & 1 & 1.6 & 0.01 & 450 & 0.15 & 0.10 & 1000 &
0.2 & 1 & 55 & 210 \\ 
IRAS 04166+2706 & B & 10 & 1 & 1 & 1.4 & 0.01 & 450 & 0.05 & 0.10 & 500 &
0.2 & 1 & 50 & 120 \\ 
IRAS 04166+2706 & C & 1 & 10 & 1 & 1.2 & 0.0 & 30 & ... & 0.05 & 500 &
0.02 & 3 & 60 & 140 \\ 
IRAS 04166+2706 & D & 1 & 1 & 10 & 1.9 & 0.01 & 450 & 0.15 & 0.10 & 2000 &
0.2 & 1 & 65 & 240 \\ 
\hline
IRAS 04169+2702 & A & 1 & 1 & 1 & 2.1 & 0.01 & 450 & 0.15 & 0.10 & 2000 &
1 & 1 & 30 & 90 \\ 
IRAS 04169+2702 & B & 10 & 1 & 1 & 1.6 & 0.01 & 250 & 0.05 & 0.10 & 500 &
0.2 & 1 & 35 & 90 \\ 
IRAS 04169+2702 & C & 1 & 10 & 1 & 1.4 & 0.01 & 450 & 0.15 & 0.10 & 2000 &
1 & 1 & 30 & 90 \\ 
IRAS 04169+2702 & D & 1 & 1 & 10 & 2.4 & 0.0 & 100 & ... & 0.05 & 500 &
0.02 & 1 & 55 & 90 \\ 
\hline
IRAS 04287+1801 & A & 0.1 & 1 & 1 & 6.6 & 0.10 & 250 & 0.05 & 0.05 & 1000
& 1 & 10 & 40 & 160 \\ 
IRAS 04287+1801 & B & 0.5 & 1 & 1 & 8.1 & 0.50 & 450 & 0.05 & 0.10 & 1000 &
0.2 & 3 & 40 & 180 \\ 
IRAS 04287+1801 & C & 0.1 & 10 & 1 & 2.8 & 0.10 & 100 & 0.05 & 0.10 & 2000
& 0.2 & 10 & 50 & 160 \\ 
IRAS 04287+1801$^{\ast}$ & D & 1 & 1 & 10 & 9.0 & 0.50 & 450 & 0.05 & 0.01 & 2000
& 0.2 & 10 & 50 & 160 \\ 
\hline
IRAS 04295+2251 & A & 1 & 1 & 1 & 3.3 & 0.01 & 100 & 0.05 & 0.05 & 1000 &
0.2 & 1 & 45 & 300 \\ 
IRAS 04295+2251 & B & 10 & 1 & 1 & 2.2 & 0.01 & 100 & 0.05 & 0.05 & 500 &
0.2 & 1 & 45 & 300 \\ 
IRAS 04295+2251 & C & 1 & 10 & 1 & 1.9 & 0.01 & 30 & 0.05 & 0.005 & 500 &
0.2 & 1 & 50 & 300  \\ 
IRAS 04295+2251 & D & 1 & 1 & 10 & 1.9 & 0.01 & 30 & 0.05 & 0.005 & 500 &
1 & 1 & 55 & 300  \\ 
\hline
IRAS 04302+2247 & A & 1 & 1 & 1 & 3.8 & 0.01 & 250 & 0.05 & 0.05 & 500 &
0.2 & 1 & 70 & 10 \\ 
IRAS 04302+2247 & B & 10 & 1 & 1 & 3.0 & 0.01 & 250 & 0.05 & 0.10 & 500 &
0.2 & 1 & 75 & 10 \\ 
IRAS 04302+2247 & C & 1 & 10 & 1 & 2.2 & 0.005 & 100 & 0.15 & 0.005 & 500
& 1 & 1 & 89 & 10 \\ 
IRAS 04302+2247 & D & 1 & 1 & 10 & 3.7 & 0.01 & 250 & 0.15 & 0.05 & 2000 &
0.2 & 1 & 70 & 10 \\ 
\hline
IRAS 04361+2547 & A & 1 & 1 & 1 & 4.8 & 0.005 & 30 & 0.15 & 0.005 & 500 &
1 & 3 & 55 & 310 \\ 
IRAS 04361+2547 & B & 10 & 1 & 1 & 2.6 & 0.01 & 450 & 0.15 & 0.01 & 500 &
1 & 3 & 45 & 280 \\ 
IRAS 04361+2547 & C & 1 & 10 & 1 & 2.1 & 0.005 & 100 & 0.15 & 0.01 & 500 &
1 & 3 & 45 & 280 \\ 
IRAS 04361+2547 & D & 1 & 1 & 10 & 5.5 & 0.005 & 30 & 0.15 & 0.005 & 500 &
1 & 3 & 55 & 310 \\  
\hline
IRAS 04365+2535 & A & 1 & 1 & 1 & 1.3 & 0.005 & 100 & 0.05 & 0.05 & 500 &
0.2 & 1 & 25 & 310 \\ 
IRAS 04365+2535 & B & 10 & 1 & 1 & 1.3 & 0.005 & 100 & 0.15 & 0.10 & 1000
& 1 & 1 & 25 & 310 \\ 
IRAS 04365+2535 & C & 1 & 10 & 1 & 1.1 &  0.005 & 100 & 0.05 & 0.05 & 500 &
0.2 & 1 & 25 & 310 \\ 
IRAS 04365+2535 & D & 1 & 1 & 10 & 1.5 &  0.005 & 100 & 0.05 & 0.05 & 500 &
0.2 & 1 & 25 & 310 \\ 
\enddata
\tablecomments{The reduced $\chi^2$ is defined by Equation
  \ref{eq:chi2}.  For each source, the first row represents our best
  attempt at providing equal weights to each dataset.  The following
  three rows show the best-fit models when one dataset is weighted up
  relative to the others.  We label each row A,B,C, or D.
In some cases, the best-fit model is the
  same for different weights, although the reduced $\chi^2$ values may
  differ.  Note that $R_{\rm disk} = R_{\rm c}$, and so this
  column is meaningful even when $M_{\rm disk}=0$. \\
$^{\ast}$---for this model we increased the 
  weight of the IRS data relative to the remaining SED data.  With
  normal weighting of the IRS data, the SED-weighted data is fit best
  with the first model listed for this object.}
\end{deluxetable}

\begin{deluxetable}{l|cccccc}
\tabletypesize{\scriptsize}
\tablewidth{0pt}
\tablecaption{Model Mass Distributions
\label{tab:massdist}}
\tablehead{\colhead{Source} & \multicolumn{5}{c}{Mass within 100 AU} &
  \colhead{Mass within 100 AU} \\
&  \multicolumn{5}{c}{(M$_{\odot}$)} &  / Total Mass\\
 & A & B & C & D & {Mean} & }
\startdata
IRAS 04016+2610 & 0.003 & 0.003 & 0.002 & 0.002 & $0.002 \pm 0.0004$ &
$0.03 \pm 0.008$ \\
IRAS 04166+2706 & 0.003 & 0.005 & 0.005 & 0.003 & $0.004 \pm 0.0009$ &
$0.05 \pm 0.03$ \\
IRAS 04169+2702 & 0.003 & 0.007 & 0.003 & 0.002 & $0.004 \pm 0.002$ &
$0.04 \pm 0.01$ \\ 
IRAS 04287+1801 & 0.05 & 0.18 & 0.09 & 0.18 & $0.12 \pm 0.06$ & $0.37
\pm 0.07$ \\
IRAS 04295+2251 & 0.01 & 0.01 & 0.01 & 0.01 & {0.01 $\pm$ 0.0008}  &
$0.44 \pm 0.30$\\
IRAS 04302+2247 & 0.006 & 0.007 & 0.005 & 0.005 & $0.006 \pm 0.001$ &
$0.18 \pm 0.18$ \\ 
IRAS 04361+2547 & 0.005 & 0.003 & 0.005 & 0.005 & $0.005 \pm 0.0008$ &
$0.40 \pm 0.17$\\
IRAS 04365+2535 & 0.007 & 0.006 & 0.007 & 0.007  & $0.007 \pm 0.0007$
& $0.11 \pm 0.04$ \\
\enddata
\tablecomments{For each source, we list the mass within 100 AU
  for the four models listed in Table \ref{tab:modfits}.  We then give
the mean value and the standard deviation.  Finally, we list the
fractional mass within 100 AU compared to the total mass.}
\end{deluxetable}

\end{document}